\documentclass[10pt,twocolumn,twoside]{IEEEtran}
\renewcommand{\baselinestretch}{0.95}

\usepackage{color}

\usepackage{times,amsmath,epsfig}
\usepackage{amssymb}
\usepackage{cite}

\usepackage{algorithmic}
\usepackage{algorithm}
\usepackage[font=small,format=plain,up,up]{caption}
\usepackage{subcaption}
\usepackage{needspace}

\usepackage{tikz}
\usetikzlibrary{shapes,arrows}
\input{mysymbol.sty}
\tikzstyle{phantom vertex} = [ ellipse, 
                               anchor = center, 
                               minimum height = 1*\unit, 
                               minimum width  = 1*\unit,
                               inner sep=0pt,
                               anchor=center]
\tikzstyle{red vertex}   = [black, fill = red!20,   phantom vertex, draw]
\tikzstyle{black vertex} = [black, fill = black!20, phantom vertex, draw]
\tikzstyle{blue vertex}  = [black, fill = blue!20,  phantom vertex, draw]
\tikzstyle{green vertex} = [black, fill = green!20,  phantom vertex, draw]
\tikzstyle{yellow vertex} = [black, fill = yellow!20,  phantom vertex, draw]
\tikzstyle{cyan vertex} = [black, fill = cyan!20,  phantom vertex, draw]
\tikzstyle{vertex}       = [draw, phantom vertex]

\tikzstyle{point} = [ellipse, inner sep=0pt, draw, fill=white, anchor = center,
                     minimum height = 0.05*\unit, minimum width  = 0.05*\unit]

\newcommand{\EE}{{\mathbb E}}

\newcommand{\mymod}{\mathrm{mod}_N}

\newtheorem{mylemma}{\hspace{-11pt}\bf Lemma}
\newtheorem{myproposition}{\hspace{-11pt}\bf Proposition}
\newtheorem{remark}{\hspace{-11pt}\bf Remark}

\psfull

\title{Sampling of graph signals with successive \\ local aggregations}
\author{\IEEEauthorblockN{Antonio G. Marques, Santiago Segarra, Geert Leus, and Alejandro Ribeiro}
\thanks{Work in this paper is supported by Spanish MINECO grant No TEC2013-
41604-R and USA NSF CCF-1217963. A. G. Marques is with the Dept. of Signal Theory and Comms., King Juan Carlos Univ.
S. Segarra and A. Ribeiro are with the Dept. of Electrical and Systems Eng., Univ. of Pennsylvania. G. Leus is with the Dept. of Electrical Eng., Mathematics
and Computer Science, Delft Univ. of Technology. Emails: antonio.garcia.marques@urjc.es, ssegarra@seas.upenn.edu, g.j.t.leus@tudelft.nl, and aribeiro@seas.upenn.edu.}}

\vspace{-1.5cm}
\begin{document}
\maketitle

\vspace{-1.5cm}

\begin{abstract}%

A new scheme to sample signals defined in the nodes of a graph is proposed. The underlying assumption is that such signals admit a sparse representation in a frequency domain related to the structure of the graph, which is captured by the so-called graph-shift operator. Most of the works that have looked at this problem have focused on using the value of the signal observed at a subset of nodes to recover the signal in the entire graph. Differently, the sampling scheme proposed here uses as input observations taken at a single node. The observations correspond to sequential applications of the graph-shift operator, which are linear combinations of the information gathered by the neighbors of the node. When the graph corresponds to a directed cycle (which is the support of time-varying signals), our method is equivalent to the classical sampling in the time domain. When the graph is more general, we show that the Vandermonde structure of the sampling matrix, which is critical to guarantee recovery when sampling time-varying signals, is preserved. Sampling and interpolation are analyzed first in the absence of noise and then noise is considered. We then study the recovery of the sampled signal when the specific set of frequencies that is active is not known. Moreover, we present a more general sampling scheme, under which, either our aggregation approach or the alternative approach of sampling a graph signal by observing the value of the signal at a subset of nodes can be both viewed as particular cases. The last part of the paper presents numerical experiments that illustrate the results developed through both synthetic graph signals and a real-world graph of the economy of the United States.
\end{abstract}


\section{Introduction}\label{S:Introduction}

Sampling (and subsequent interpolation) is a cornerstone problem in classical signal processing \cite{unser2000sampling}. The emergence of new fields of knowledge such as network science and big data is generating a pressing need to extend the results existing for classical time-varying signals to signals defined on graphs \cite{EmergingFieldGSP,SandryMouraSPG_TSP13,RabICASSP12_ApproxSignalsGraphs}. This not only entails modifying the algorithms currently available for time-varying signals, but also gaining intuition on what concepts are preserved (and lost) when a signal is  defined, not in the classical time grid, but in a more general graph domain.

This paper investigates the sampling and posterior recovery of signals that are defined in the nodes of a graph. The underlying assumption is that such signals admit a sparse representation in a (frequency) domain which is related to the structure of the graph where these signals reside. Most of the current efforts in this field have been focused on using the value of the signal observed at a subset of nodes to recover the signal in the entire graph \cite{SamplingOrtegaICASSP14,AlgFindSupportSamplGlobalsip2014,SamplingKovacevicMoura_1415,chen2015discrete,wang2014local}. Our proposal in this paper is different. We present a new sampling method that accounts for the graph structure, can be run at a single node and only requires access to information of neighboring nodes. Moreover, we also show that the proposed method shares similarities with the classical sampling and interpolation of time-varying signals. When the graph corresponds to a directed cycle, which is the support of classical time-varying signals, our method is equivalent to classical sampling. When the graph is more general, the Vandermonde structure of the sampling matrix, which is critical to guarantee recovery in classical sampling \cite{unser2000sampling}, is preserved. Such a structure not only facilitates the interpolation process, but also helps to draw some connections between the proposed method and the sampling of time-varying signals. Sampling and interpolation are analyzed first in the absence of noise, where the conditions under which recovery is guaranteed are identified. The conditions depend both on the structure of the graph and the particular node taking the observations. They also reveal that one way to understand bandlimited graph signals is to think of signals that can be well approximated by only observing the value of the signal at a small neighborhood. We then analyze the sampling and reconstruction process when noise is present and when the specific frequencies where the signal is sparse are not known. For the noisy case, an interpolator based on the Best Linear Unbiased Estimator (BLUE) is designed and the effect on the corresponding error covariance matrix of different noise models is discussed. For the case of unknown frequency support, we also provide conditions under which the signal can be identified. This second problem falls into the category of sparse signal reconstruction \cite{donoho2003spark,candes2006robust,elad2007optimized,RabICASSP12_SpectCompressSensingGraphs} where the main idea is to leverage the structure of the observation matrix to facilitate recovery. The last contribution is the design of a generalization of our sampling method that considers a subset of nodes, each of them taking multiple observations. Within that generalization, the approach of sampling a graph signal by observing the value of the signal at a subset of nodes can be viewed as a particular case. Hence, the generalization will also be useful to compare and establish relationships between existing approaches to sample signals in graphs and our proposed method.

The paper is organized as follows. Section~\ref{S:Modeling} introduces the new aggregation sampling method, compares it to the existing selection sampling method and shows that for classical time-varying signals both methods are equivalent. Section~\ref{S:SamplingBandlimited} analyzes our sampling method in more detail and applies it to sample bandlimited graph signals. The analysis includes conditions for recovery, which are formally stated in Section~\ref{S:sampling_GS_local_aggregat}. Section~\ref{S:Noisy_samp_recov} investigates the effect of noise in aggregation sampling. It also discusses how to select sampling nodes and observation schemes that lead to a good recovery performance. Corresponding modifications in the interpolation in order to recover the signal when the support is not known are discussed in Section~\ref{S:unknown_support}. Section~\ref{S:shift_space_bigvec_sampling} proposes a generalization under which the existing selection sampling and the proposed aggregation sampling can be viewed as particular cases. Several illustrative numerical results are presented in Section~\ref{S:NumExper}. A few concluding remarks are provided in Section~\ref{S:Concl}, which closes the paper.

\noindent \textbf{Notation:} Boldface capital letters denote matrices and boldface lowercase letters column vectors. Generically, the entries of a matrix $\mathbf{X}$ and a vector $\mathbf{x}$ are denoted as $X_{ij}$ and $x_i$; however, when contributing to avoid confusion, the alternative notation $[\mathbf{X}]_{ij}$ and $[\mathbf{x}]_i$ will be used. The notations $^T$ and $^H$ stand for transpose and transpose conjugate, respectively; $\otimes$ is the Kronecker product; $\mathrm{trace}(\mathbf{X}):=\sum_{i} X_{ii}$ is the trace of the square matrix $\mathbf{X}$ and $\mathrm{det}(\mathbf{X})$ is its determinant; $\diag(\mathbf{x})$ is a diagonal matrix satisfying $[\diag(\mathbf{x})]_{ii}=[\mathbf{x}]_i$; $\mathrm{vec}(\mathbf{X})$ is the column-wise vectorized version of matrix $\mathbf{X}$; $\mathbf{e}_i$ is the $i$-th $N\times 1$ canonical vector (all entries of $\mathbf{e}_i$ are zero except the $i$-th one, which is one); $\mathbf{E}_K:=[\mathbf{e}_1,...,\mathbf{e}_K]$ is a tall matrix collecting the $K$ first canonical vectors; and $\mathbf{0}$ and $\mathbf{1}$ are, respectively, the all-zeros and all-ones matrices (when not clear from the context, a subscript indicating the dimensions will be used). The modulus (remainder) obtained after dividing $x$ by $N$ is denoted as $\mymod(x)$.

\section{Sampling of graph signals}\label{S:Modeling}

Let $\ccalG=(\ccalN,\ccalE)$ denote a directed graph. The set of nodes $\ccalN$ has cardinality $N$, and the set of links $\ccalE$ is such that  $(i,j)\in\ccalE$ if and only if node $i$ is connected to node $j$. The set $\ccalN_i:\{\, j \, |(j,i)\in\ccalE\}$ contains all nodes with an incoming connection to $i$ and is termed the incoming neighborhood of $i$. For any given graph we define the adjacency matrix $\bbA$ as a sparse $N\times N$ matrix with nonzero elements $A_{ji}$ if and only if $(i,j)\in\ccalE$. The value of $A_{ji}$ captures the strength of the connection between $i$ and $j$. When the graph is unweighted, the nonzero elements of $\mathbf{A}$ are set to one. The focus of this paper is not on analyzing $\ccalG$, but a graph signal defined on the set of nodes $\ccalN$. Such a signal can be represented as a vector $\bbx=[x_1,\ldots,x_N]^T \in  \mathbb{R}^N$ where the $i$-th component represents the value of the signal at node $i$, or, equivalently, as a function $f : \ccalN \to \mathbb{R}$, defined on the vertices of the graph.


The graph $\ccalG$ is endowed with a \emph{graph-shift operator} $\bbS$ defined as an $N\times N$ matrix whose entry $(i,j)$, denoted as $S_{ij}$, can be nonzero only if $i=j$ or $(j,i)\in\ccalE$. The sparsity pattern of the matrix $\bbS$ captures the local structure of $\ccalG$ but we make no specific assumptions on the values of the nonzero entries of $\bbS$. Common choices for $\bbS$ are the adjacency matrix of the graph \cite{SandryMouraSPG_TSP13,SandryMouraSPG_TSP14Freq}, the Laplacian \cite{EmergingFieldGSP}, and its generalizations \cite{godsil2001algebraic}. The intuitive interpretation of $\bbS$ is that it represents a linear transformation that can be computed locally at the nodes of the graph. If $\bby=[y_1,\ldots,y_N]^T$ is defined as $\bby=\bbS\bbx$, then node $i$ can compute $y_i$ provided that it has access to the values of $x_j$ at its incoming neighbors $j\in \ccalN_i$. We assume henceforth that $\bbS$ is diagonalizable, so that there exists a $N\times N$ matrix $\bbV$ and a $N\times N$ diagonal matrix $\bbLambda$ that can be used to decompose $\bbS$ as
\begin{equation}\label{eqn_eigendecomposition}
   \bbS=\bbV\bbLambda\bbV^{-1}.
\end{equation}
In particular, \eqref{eqn_eigendecomposition} is true for normal matrices satisfying $\bbS\bbS^H=\bbS^H\bbS$. In that case we have that $\bbV$ is unitary, which implies $\bbV^{-1}=\bbV^{H}$, and leads to the decomposition $\bbS=\bbV\bbLambda\bbV^H$.

%
\begin{figure}
\centering
\input{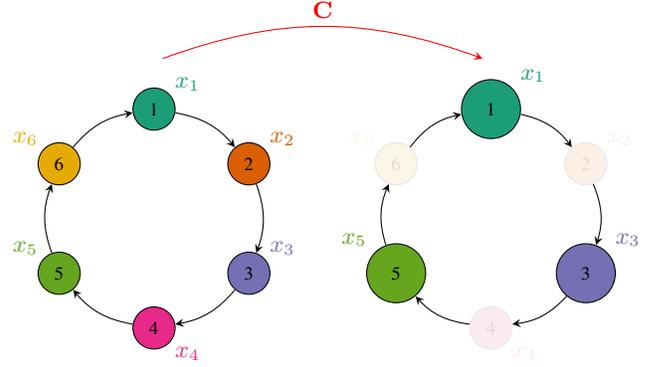}
\def \thisplotscale {1.12}

\def \unit {\thisplotscale cm}

\tikzstyle {vertex} = [circle, 
                       draw,
                       minimum width = 0.5*\unit,
                       minimum height = 0.5*\unit,
                       anchor=center,
                       font=\scriptsize]
\tikzstyle {light}  = [opacity = 0.1]
\tikzstyle{bigvertex} = [vertex, 
                        minimum width  = 0.7*\unit,
                        minimum height = 0.7*\unit]

{\small \begin{tikzpicture}[x = 1*\unit, y = 1*\unit]
    
	\node at (0, 0) (center1) {};
	\path (center1) ++ (  90:1.3) node (1) [fill = lightred,    vertex] {1};
	\path (center1) ++ (  30:1.3) node (2) [fill = lightorange, vertex] {2};
	\path (center1) ++ (- 30:1.3) node (3) [fill = lightyellow,  vertex] {3};
	\path (center1) ++ (- 90:1.3) node (4) [fill = lightgreen,  vertex] {4};
	\path (center1) ++ (-150:1.3) node (5) [fill = lightblue,   vertex] {5};
	\path (center1) ++ ( 150:1.3) node (6) [fill = lightpurple, vertex] {6};
	\path (1) ++ ( 0.4, 0.3) node [darkred   ] {$x_1$};
	\path (2) ++ ( 0.4, 0.3) node [darkorange] {$x_2$};
	\path (3) ++ ( 0.4, 0.3) node [darkyellow] {$x_3$};
	\path (4) ++ ( 0.4,-0.3) node [darkgreen ] {$x_4$};
	\path (5) ++ (-0.4, 0.3) node [darkblue  ] {$x_5$};
	\path (6) ++ (-0.4, 0.3) node [darkpurple] {$x_6$};
	\path[-stealth] (1) edge [bend left=20, above] node {} (2);		
	\path[-stealth] (2) edge [bend left=20, above] node {} (3);		
	\path[-stealth] (3) edge [bend left=20, above] node {} (4);		
	\path[-stealth] (4) edge [bend left=20, above] node {} (5);		
	\path[-stealth] (5) edge [bend left=20, above] node {} (6);
	\path[-stealth] (6) edge [bend left=20, above] node {} (1);	

	\node at (4, 0) (center2) {};
	\path (center2) ++ (  90:1.3) node (1) [fill = lightred,           bigvertex] {1};
	\path (center2) ++ (  30:1.3) node (2) [fill = lightorange, light, vertex] {2};
	\path (center2) ++ (- 30:1.3) node (3) [fill = lightyellow,        bigvertex] {3};
	\path (center2) ++ (- 90:1.3) node (4) [fill = lightgreen,  light, vertex] {4};
	\path (center2) ++ (-150:1.3) node (5) [fill = lightblue,          bigvertex] {5};
	\path (center2) ++ ( 150:1.3) node (6) [fill = lightpurple, light, vertex] {6};
	\path (1) ++ ( 0.5, 0.4) node [darkred          ] {$x_1$};
	\path (2) ++ ( 0.4, 0.3) node [darkorange, light] {$x_2$};
	\path (3) ++ ( 0.5, 0.4) node [darkyellow       ] {$x_3$};
	\path (4) ++ ( 0.4,-0.3) node [darkgreen, light ] {$x_4$};
	\path (5) ++ (-0.5, 0.4) node [darkblue         ] {$x_5$};
	\path (6) ++ (-0.4, 0.3) node [darkpurple, light] {$x_6$};
	\path[-stealth] (1) edge [bend left=20, above] node {} (2);		
	\path[-stealth] (2) edge [bend left=20, above] node {} (3);		
	\path[-stealth] (3) edge [bend left=20, above] node {} (4);		
	\path[-stealth] (4) edge [bend left=20, above] node {} (5);		
	\path[-stealth] (5) edge [bend left=20, above] node {} (6);
	\path[-stealth] (6) edge [bend left=20, above] node {} (1);	

    \path (0,1.9) node (top) {};
    \path[-stealth] (center1.east |- top) 
                    edge [bend left=20, above, red] node 
                    {$\bbC$} 
                    (center2.west |- top);

\end{tikzpicture}} 
\caption{Sampling in the time domain as selection sampling on a directed cycle graph. The discrete time domain can be represented by a cyclic graph connecting node $i$ to $i+1$ and node $N$ to node 1. Using the uniform selection matrix $\bbC=[\bbe_1,\bbe_{N/K+1},\ldots,\bbe_{N-N/K+1}]^T$ with $K/N=1/2$ results in the selection of the signal values at odd indexed nodes $x_1,x_3,\ldots,x_{N-1}$. Observe that this sampling rule is independent of the structure of the underlying graph.}
\label{fig_selection_sampling}
\end{figure}

%
A natural definition of sampling for a graph signal is to introduce a fat $K\times N$ selection
matrix $\bbC$ and define the sampled signal as \cite{chen2015discrete}
\begin{equation}\label{eqn_selection_sampling}
   \bar{\bbx} = \bbC\bbx.
\end{equation}
If the matrix $\bbC$ is chosen as binary, i.e., with elements $C_{ij}\in\{0,1\}$, has a single nonzero element per row, and at most one nonzero element per column, then the signal $\bar{\bbx}$ is a selection of $K$ out of the $N$ elements of $\bbx$. In such a case, the ratio $K/N$ is the sampling rate of the signal. Uniform sampling amounts to setting $\bbC=[\bbe_1,\bbe_{N/K+1},\ldots,\bbe_{N-N/K+1}]^T$ and the selection of the first $K$ elements of $\bbx$ is accomplished by setting $\bbC=\bbE^T_K:=[\bbe_1,\ldots,\bbe_{K}]^T$. We remark that, in general, it is not clear how to choose good selection matrices $\bbC$. This is in contrast to conventional sampling of signals in the time domain where uniform sampling is advantageous.

%
\begin{figure*}
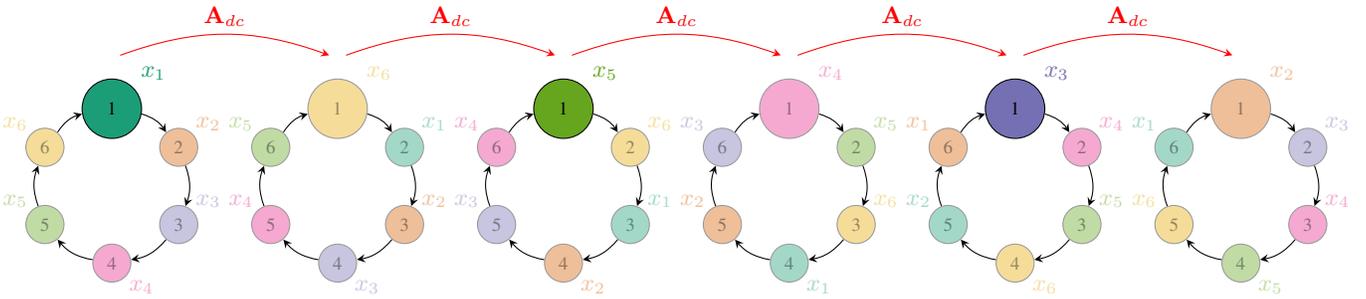

\centering
\input{figures/graph_palette}
\input{figures/cyclic_sampling.tex}
\caption{Sampling in the time domain as aggregation sampling in a directed cycle graph. In aggregation sampling we utilize successive applications of a shift operator determined by the given graph and sample the resulting signal observed at a given node. Using the cycle adjacency matrix $\bbA_{dc}$ as shift operator results in the signal $\bbx$ rotating through the graph and the selection of elements of the aggregated signal reduces to conventional sampling. Aggregation sampling is, therefore, a generalization of conventional sampling to graph signals that utilizes the underlying graph structure in the construction of samples.}
\vspace{-0.1in}
\label{F:DirectedChainGraph_sampling}
\end{figure*}

An equally valid, yet less intuitive, definition is to fix a node, say $i$, and consider the sampling of the signal seen by this node as the shift operator $\bbS$ is applied recursively. To describe this sampling methodology more clearly, define the $l$-th shifted signal $\bby^{(l)}:=\bbS^l\bbx$ and further define the $N\times N$ matrix
\begin{equation}\label{eqn_all_shifts_matrix}
   \bbY\,:=\,[\bby^{(0)},\bby^{(1)},\ldots,\bby^{(N-1)}]
       \,=\,[\bbx,\bbS\bbx,\ldots,\bbS^{N-1}\bbx],
\end{equation}
that groups the signal $\bbx$ and the result of the first $N-1$ applications of the shift operator. Associating the $i$-th row of $\bbY$ with node $i$, we define the successively aggregated signal at $i$ as $\bby_i:=(\bbe_i^T\bbY)^T=\bbY^T\bbe_i$. Sampling is now reduced to the selection of $K$ out of the $N$ elements (rows) of $\bby_i$, which we accomplish with a selection matrix $\bbC$ [cf. \eqref{eqn_selection_sampling}]
\begin{equation}\label{eqn_aggregation_sampling}
   \bar{\bby}_i \ :=\ \bbC\bby_i\ =\ \bbC \left(\bbY^T\bbe_i\right).
\end{equation}
We say that the signal $\bar{\bby}_i$ samples $\bbx$ with successive local aggregations. This nomenclature follows from the fact that $\bby^{(l)}$ can be computed recursively as $\bby^{(l)}:=\bbS\bby^{(l-1)}$ and that the $i$-th element of this vector can be computed using signals associated with itself and its incoming neighbors,
\begin{equation}\label{eqn_local_aggregation}
   y^{(l)}_i = \sum_{j\in\ccalN_i} S_{ij} y^{(l-1)}_j.
\end{equation}
We can then think of the signal $\bby_i$ as being computed locally at node $i$ using successive variable exchanges with neighboring nodes. In fact, it is easy to show that $y_i^{(l)}$ can be expressed as a linear combination of the values of $x_j$ at nodes $j$ whose distance (number of hops) from node $i$ is less than or equal to $l$. This implies that the sampled signal $\bar{\bby}_i$ in \eqref{eqn_aggregation_sampling} is a selection of values that node $i$ can determine locally. An underlying idea behind the sampling in \eqref{eqn_aggregation_sampling} is to incorporate the structure of the shift into the sampling procedure. Indeed, $\mathbf{S}$ and  $\mathbf{y}^{(l)}$ play key roles in other graph-processing algorithms such as shift-invariant graph filters \cite{SandryMouraSPG_TSP13}, where the output of the filter can be viewed a linear combination of the shifted signals $\mathbf{y}^{(l)}$.

To understand the difference between selection sampling [cf. \eqref{eqn_selection_sampling}] and aggregation sampling [cf. \eqref{eqn_aggregation_sampling}], it is instructive to consider their application to a signal defined in the time domain. We do so in the following section.

%
\subsection{Example: Sampling in the time domain}

Classical time domain signals can be represented as graph signals defined on top of a directed cycle graph \cite{EmergingFieldGSP,chen2015discrete}, as illustrated in Fig.~\ref{fig_selection_sampling}. To do so define the directed cycle graph $\ccalG_{dc}$ as one in which the edge set $\ccalE_{dc}:=\{(i,\mymod(i)+1)\}_{i=1}^N$, connects node $i$ to node $i+1$ for all nodes except $N$, which is connected to node 1. The elements of the adjacency matrix of this graph, denoted as $\bbA_{dc}$, are zero except for the ones in the first cyclic subdiagonal, which are one.

For a signal $\bbx$ defined on top of the directed cycle $\ccalG_{dc}$, we consider selection sampling and aggregation sampling when using the shift operator $\bbS=\bbA_{dc}$ and the uniform selection matrix $\bbC=[\bbe_1,\bbe_{N/K+1},\ldots,\bbe_{N-N/K+1}]^T$. Illustrations of the respective sampling procedures are available in Figs.~\ref{fig_selection_sampling} and \ref{F:DirectedChainGraph_sampling} for a signal with $N=6$ elements and sampling rate $K/N=1/2$.

In selection sampling we just multiply the graph signal $\bbx$ with the selection matrix $\bbC$ to obtain the sampled signal $\bar{\bbx} = \bbC\bbx$ as indicated by \eqref{eqn_selection_sampling}. In aggregation sampling we consider subsequent applications of the shift matrix $\bbS=\bbA_{dc}$. Each of these shifts amounts to rotating the signal clockwise so that the element at node $i$ moves to node $i+1$ for all $i<N$ and the element at node $N$ moves to node $1$. If we consider, e.g., node $i=1$, the first shift moves signal $x_{N}$ to this node so that $y_{1}^{(1)}=x_N$, the second shift moves signal $x_{N-1}$ to this node so that $y_{1}^{(2)}=x_{N-1}$ and so on. It follows that the aggregated signal $\bby_1$ in \eqref{eqn_all_shifts_matrix} is given by $\bby_1=[x_1, x_{N}, x_{N-1},\ldots,x_2]$. This is just a shift of the original signal $\bbx$, which, upon multiplication by the selection matrix $\bbC$ as per \eqref{eqn_aggregation_sampling} results in a vector $\bar{\bby}_1=\bbC\bby_1$ that contains the same elements that $\bar{\bbx}$ contains.

For the cycle graph and shift operator $\bbS=\bbA_{dc}$ selection and aggregation sampling produce not only equivalent sampled signals but also reduce to conventional sampling. This is not a coincidence because both methods are designed as generalizations of conventional sampling. In general, selection sampling and aggregation sampling produce different outcomes. In selection sampling we move through nodes to collect samples at points uniquely identified by $\bbC$, whereas in aggregation sampling we move the signal through the graph while collecting samples at a fixed node. Observe that because aggregation sampling depends on the shift operator, it incorporates the structure of the graph into the sampling procedure. This is not true for selection sampling except for the choice of matrices $\bbC$ adapted to particular graphs.

\section{Sampling of bandlimited graph signals}\label{S:SamplingBandlimited}

Recovery of the original signal from its sampled version is possible under the assumption that the original signal admits a sparse representation. This section begins by introducing the concept of a bandlimited graph signal, which is sparse in the frequency domain, and establishing some connections with the concept of bandlimitedness in the classical time domain. Section~\ref{Ss:conventional sampling} reviews briefly the recovery of a bandlimited graph signal for the case of selection sampling. Section~\ref{S:sampling_GS_local_aggregat} analyzes the recovery of a bandlimited graph signal for the case of aggregation sampling.

\subsection{Bandlimited graph signals}\label{SS:BandlimitedGraphSig}

The common practice when addressing the problem of sampling signals in graphs is to suppose that the graph-shift operator $\mathbf{S}$ plays a key \emph{role in explaining} the signals of interest $\mathbf{x}$. More specifically, that $\mathbf{x}$ can be expressed as a linear combination of a \emph{subset} of the columns of $\mathbf{V}=[\mathbf{v}_1,...,\mathbf{v}_N]$, or, equivalently, that the vector $\widehat{\mathbf{x}}=\mathbf{V}^{-1}\mathbf{x}$ is sparse. In this context, vectors $\mathbf{v}_k$ are interpreted as the graph frequency basis and $\widehat{x}_k$ as the corresponding signal frequency coefficients. To simplify exposition, it will be assumed throughout the paper that the active frequencies are the first $K$ ones, which are associated with the largest eigenvalues \cite{SamplingOrtegaICASSP14,RabICASSP14_SpectCharSigsSmallWord}. Under this assumption, it holds that $\widehat{\mathbf{x}}=[\widehat{x}_1,...,\widehat{x}_K,0,...,0]^T$. However, the results presented in the paper can be applied to any set of active frequencies $\mathcal{K}$ of size $K$ provided that $\mathcal{K}$ is known. For convenience, we define $\mathbf{V}_K:=[\mathbf{v}_1,...,\mathbf{v}_K]$ and $\widehat{\mathbf{x}}_K:=[\widehat{x}_1,...,\widehat{x}_K]^T$ so that we may write $\widehat{\mathbf{x}}=[\widehat{\mathbf{x}}^T_K  ~| ~  \mathbf{0}_{1 \times N-K}]^T$. For $\widehat{\mathbf{x}}$ to be sparse, it is reasonable to assume that $\mathbf{S}$ is involved in the generation of $\mathbf{x}$.

When $\mathcal{G}=\mathcal{G}_{dc}$, it can be easily shown that setting the shift operator either to $\mathbf{S}=\mathbf{A}_{dc}$ or to $\mathbf{S}=\mathbf{L}_{dc}:=\mathbf{I}-\mathbf{A}_{dc}$ gives rise to the Fourier basis $\mathbf{F}$. More formally, that the right eigenvectors of $\mathbf{S}$ satisfy $\mathbf{V}=\mathbf{F}$, with $F_{ij}:=\frac{1}{\sqrt{N}}e^{+\mathfrak{j}\frac{2\pi}{N}(i-1)(j-1)}$ and $\mathfrak{j}:=\sqrt{-1}$. Selecting $\mathbf{S}=\mathbf{A}_{dc}$ has the additional advantage of satisfying $\Lambda_{ii}=e^{-\mathfrak{j}\frac{2\pi}{N}(i-1)}$, i.e., the eigenvalues of the shift operator correspond to the classical discrete frequencies. Interpretations for the eigenvalues of the Laplacian matrix $\mathbf{L}_{dc}$ also exist \cite{EmergingFieldGSP}.

\subsection{Selection sampling of bandlimited graph signals}\label{Ss:conventional sampling}

Under the selection sampling approach \cite{SamplingOrtegaICASSP14,AlgFindSupportSamplGlobalsip2014,wang2014local,SamplingKovacevicMoura_1415,chen2015discrete}, sampling a graph signal amounts to setting $\bar{\mathbf{x}}=\mathbf{C}\mathbf{x}$ [cf. \eqref{eqn_selection_sampling}]. Since the $K\times N$ binary selection matrix $\mathbf{C}$ indexes the nodes that are observed, the issue then is how to design $\mathbf{C}$, i.e., which nodes to select, and how to recover the original signal $\mathbf{x}$ from its samples $\bar{\mathbf{x}}$.

To answer these questions, it is assumed that the signal $\mathbf{x}$ is bandlimited, so that it can be expressed as a linear combination of the $K$ principal eigenvectors in $\mathbf{V}$. The sampled signal $\bar{\mathbf{x}}$ is then $\bar{\mathbf{x}}=\mathbf{C}\mathbf{x}=\mathbf{C}\mathbf{V}_K\widehat{\mathbf{x}}_K$. Clearly, if the matrix $\mathbf{C}\mathbf{V}_K$ is invertible, then $\widehat{\mathbf{x}}_K$ can be recovered from $\bar{\mathbf{x}}$. Once the coefficients $\widehat{\mathbf{x}}_K$ are known, the signal in the original domain can be found as $\mathbf{x}=\mathbf{V}_K\widehat{\mathbf{x}}_K$. Combining the previous equations, we have
\begin{equation}\label{E:interp_regsampling}
\mathbf{x}=\mathbf{V}_K\widehat{\mathbf{x}}_K=\mathbf{V}_K(\mathbf{C}\mathbf{V}_K)^{-1}\bar{\mathbf{x}}.
\end{equation}
The expression in \eqref{E:interp_regsampling} shows how the original signal can be interpolated from its samples. For the previous equation to hold true, the matrix $\mathbf{C}\mathbf{V}_K$ has to be invertible. Hence, the key for guaranteeing perfect signal reconstruction is to select a subset of nodes such that the corresponding rows in $\mathbf{V}_K$ are linearly independent. In the classical domain of time-varying signals, the (Fourier) basis has a Vandermonde structure, both row-wise and column-wise. This readily implies that any subset of $K$ rows will give rise to a (row-wise) Vandermonde matrix and, hence, invertibility is guaranteed. However, for an arbitrary graph this is not guaranteed and algorithms to select a specific subset that guarantees recovery are required \cite{AlgFindSupportSamplGlobalsip2014}. The role of the Vandermonde structure of the sampling matrix will be analyzed in more detail in the ensuing sections.

\subsection{Aggregation sampling of bandlimited graph signals}\label{S:sampling_GS_local_aggregat}

As explained in \eqref{eqn_aggregation_sampling}, under the aggregation approach the sampled signal is formed by observations of the shifted signals $\mathbf{y}^{(l)}=\mathbf{S}^l\mathbf{x}$ taken at a given node $i$. Under this second approach, the graph-shift operator $\mathbf{S}$ plays a key \emph{role} not only in \emph{explaining} and \emph{recovering} $\mathbf{x}$, but also in \emph{sampling} $\mathbf{x}$. Another reason to consider this scheme is that the entries of $\mathbf{y}^{(l)}$ can be found by sequentially exchanging information among neighbors. This implies that: a) for setups where graph vertices correspond to nodes of an actual network, the procedure can be implemented distributedly; and b) if recovery is feasible, the observations at a single node can be used to recover the signal in the entire graph.

Mimicking the approach in the previous section, we first analyze how the bandlimitedness of $\bbx$ is manifested on the sampled signal. Then, we identify under which conditions recovery is feasible and describe the corresponding interpolation algorithm. For the ease of exposition, the dependence of $\mathbf{y}_i$ on $\widehat{\mathbf{x}}$ is given in the form of a lemma.

\begin{mylemma}\label{L:vandermonde_matrix}
\emph{Define the $N\times 1$ vector $\boldsymbol{\upsilon}_i:=\mathbf{V}^T\mathbf{e}_i$, which collects the values of the frequency basis $\{\mathbf{v}_k\}_{k=1}^K$ at node $i$, and the $N\times N$ (column-wise) Vandermonde matrix
\begin{equation}
\boldsymbol{\Psi}:= \left( \begin{array}{ccc}
1 &  \ldots & 1 \\
\lambda_1 & \ldots & \lambda_N\\
\vdots &  &\vdots\\
\lambda_1^{N-1} &  \ldots  & \lambda_N^{N-1} \end{array} \right).
\end{equation}
Then, the shifted signal $\mathbf{y}_i$ can be expressed as
\begin{equation}\label{E:shifted_signal}
\mathbf{y}_i=\boldsymbol{\Psi}\diag(\boldsymbol{\upsilon}_i)\widehat{\mathbf{x}}.
\end{equation}}
\end{mylemma}
\begin{myproof}
Using the spectral decomposition of $\mathbf{S}$, signal $\mathbf{y}^{(l)}$ can be written as
\begin{equation}\label{E:def_shifted_vec}
     \mathbf{y}^{(l)}=\mathbf{S}^l\mathbf{x}=(\mathbf{V}\boldsymbol{\Lambda}^l\mathbf{V}^{-1})\mathbf{x}=(\mathbf{V}\boldsymbol{\Lambda}^l)\widehat{\mathbf{x}}.
     \end{equation}

Based on the definitions of $\mathbf{y}_i$ and $\boldsymbol{\upsilon}_i$, it follows that
\begin{eqnarray}
\nonumber \mathbf{y}_i&=&\mathbf{Y}^T\mathbf{e}_i=(\mathbf{V}\mathbf{V}^{-1}\mathbf{Y})^T\mathbf{e}_i\\
&=&(\mathbf{V}^{-1}\mathbf{Y})^T\mathbf{V}^T\mathbf{e}_i=(\mathbf{V}^{-1}\mathbf{Y})^T\boldsymbol{\upsilon}_i.
\end{eqnarray}
Since the $l$-th column of matrix $\mathbf{Y}$ is $\mathbf{y}^{(l-1)}$, it can be written as $(\mathbf{V}\boldsymbol{\Lambda}^{l-1})\widehat{\mathbf{x}}$ [cf. \eqref{E:def_shifted_vec}]. Hence, the $l$-th column of matrix $(\mathbf{V}^{-1}\mathbf{Y})$ can be written as $\boldsymbol{\Lambda}^{l-1}\widehat{\mathbf{x}}$ or, equivalently, as $\diag(\widehat{\mathbf{x}})[\lambda_1^{l-1},...,\lambda_N^{l-1}]^T$.
Leveraging the fact that the vector containing the $l$-th power of the eigenvalues corresponds to the row $l+1$ of matrix $\boldsymbol{\Psi}$, the shifted signal $\mathbf{y}_i$ can be expressed as
\begin{eqnarray}\label{E:shifted_signal_proof}
\nonumber \mathbf{y}_i&=&(\mathbf{V}^{-1}\mathbf{Y})^T\boldsymbol{\upsilon}_i=(\diag(\widehat{\mathbf{x}})\boldsymbol{\Psi}^T)^T\boldsymbol{\upsilon}_i\\
&=&\boldsymbol{\Psi}\diag(\widehat{\mathbf{x}})\boldsymbol{\upsilon}_i=\boldsymbol{\Psi}\diag(\boldsymbol{\upsilon}_i)\widehat{\mathbf{x}},
\end{eqnarray}
which is the claim in the lemma.
\end{myproof}

Notice that while in Section~\ref{Ss:conventional sampling} the relationship between the sparse frequency coefficients $\widehat{\mathbf{x}}$ and the signal to be sampled was simply given by $\mathbf{x}=\mathbf{V}\widehat{\mathbf{x}}$, now it is given by $\mathbf{y}_i=\boldsymbol{\Psi}\diag(\boldsymbol{\upsilon}_i)\widehat{\mathbf{x}}$.

Next, we use Lemma~\ref{L:vandermonde_matrix} to identify under which conditions recovery is feasible. To do this, let us define the $N\times K$ matrix $\boldsymbol{\Psi}_i=\boldsymbol{\Psi}\diag(\boldsymbol{\upsilon}_i)\mathbf{E}_K$. Then, the sampled signal $\bar{\mathbf{y}}_i$ is
\begin{equation}\label{E:sampled_shifted_signal}
\bar{\mathbf{y}}_i=\mathbf{C}\mathbf{y}_i=\mathbf{C}\boldsymbol{\Psi}\diag(\boldsymbol{\upsilon}_i)\widehat{\mathbf{x}}=\mathbf{C}\boldsymbol{\Psi}_i\widehat{\mathbf{x}}_K,
\end{equation}
where $\mathbf{C}$ is the binary $K\times N$ selection matrix, and $\widehat{\mathbf{x}}_K$ the vector collecting the non-zero components of $\widehat{\mathbf{x}}$. To simplify exposition, for the time being we will assume that $\mathbf{C}\!=\!\mathbf{E}_K^T$, i.e., that the observations correspond to the original signal and the first $K\!-\!1$ shifts. This assumption can be relaxed, as discussed in Remark~\ref{rem_structure_observation_matrix}.

If matrix $\mathbf{C}\boldsymbol{\Psi}_i$ is invertible, then $\widehat{\mathbf{x}}_K$ can be recovered from $\bar{\mathbf{y}}_i$ [cf. \eqref{E:sampled_shifted_signal}] and, once $\widehat{\mathbf{x}}_K$ is known, $\mathbf{x}$ can be found as $\mathbf{x}=\mathbf{V}_K\widehat{\mathbf{x}}_K$. Combining the previous expressions, we have [cf. \eqref{E:interp_regsampling}]
\begin{equation}\label{E:interp_oursampling}
\mathbf{x}=\mathbf{V}_K\widehat{\mathbf{x}}_K=\mathbf{V}_K(\mathbf{C}\boldsymbol{\Psi}_i)^{-1}\bar{\mathbf{y}}_i.
\end{equation}
The expression in \eqref{E:interp_oursampling} shows how the original signal can be interpolated from its samples. As already stressed, for the previous equation to hold true, the matrix $\mathbf{C}\boldsymbol{\Psi}_i$ has to be invertible. Hence, the key for guaranteeing perfect signal reconstruction is to select samples such that the corresponding rows in $\boldsymbol{\Psi}_i$ are linearly independent. While for the selection sampling described in Section~\ref{Ss:conventional sampling} there is no straightforward way to check the invertibility of $\mathbf{C}\mathbf{V}_K$ (existing algorithms typically do that by inspection \cite{AlgFindSupportSamplGlobalsip2014}), for the aggregation sampling described in \eqref{E:shifted_signal}-\eqref{E:interp_oursampling} the invertibility of $\mathbf{C}\boldsymbol{\Psi}_i$ can be guaranteed if the conditions presented in the following proposition hold.

\begin{myproposition}\label{P:recoveringcond_oursampling}
\emph{Let $\mathbf{x}$ and $\bar{\mathbf{y}}_i$ be, respectively, a bandlimited graph signal with at most $K$ non-zero frequency components and the output of the sampling process defined in \eqref{E:sampled_shifted_signal}. Then, the $N$ entries of signal $\mathbf{x}$ can be recovered from the $K$ samples in $\bar{\mathbf{y}}_i$ if the two following conditions hold\\
\noindent i) The first $K$ eigenvalues of the graph-shift operator $\mathbf{S}$ are distinct; i.e., $\lambda_i\neq\lambda_j$ for all $i\neq j$, $i\leq K$ and $j\leq K$.\\
\noindent ii) The $K$ first entries of $\boldsymbol{\upsilon}_i$ are non-zero.}
\end{myproposition}
\begin{myproof}
To prove the proposition it suffices to show that \emph{i)} and \emph{ii)} guarantee the invertibility of $\mathbf{C}\boldsymbol{\Psi}_i$ [cf. \eqref{E:interp_oursampling}]. Matrix $\mathbf{C}\boldsymbol{\Psi}_i$ can be understood as the multiplication of two matrices: matrix $(\mathbf{C}\boldsymbol{\Psi} \mathbf{E}_K)$ and matrix $(\mathbf{E}_K^T \diag(\boldsymbol{\upsilon}_i)\mathbf{E}_K)$. It is immediate that condition \emph{ii)} guarantees that the second matrix is invertible. Moreover, condition \emph{i)} guarantees invertibility of the first matrix. To see this, note that $(\boldsymbol{\Psi} \mathbf{E}_K)$  is a $N\times K$ (column-wise) Vandermonde matrix. Hence $\mathbf{C}(\boldsymbol{\Psi} \mathbf{E}_K)$ is a selection of the first $K$ rows of $(\boldsymbol{\Psi} \mathbf{E}_K)$, which is also Vandermonde. Any square Vandermonde matrix has full rank provided that the basis (i.e., the eigenvalues of $\mathbf{S}$) are distinct, as required in condition \emph{i)}.
\end{myproof}

\noindent One of the implications of the proposition is that there is no need to compute or observe the entire vector $\mathbf{y}_i$, since its first $K$ entries suffice to guarantee recovery.

The conditions in Proposition~\ref{P:recoveringcond_oursampling} are not difficult to check and they provide additional insights on the behavior of the sampling and interpolation procedure. Condition \emph{i)} refers to the structure of the entire graph. It states that if a graph has two identical frequencies and the signal of interest is a linear combination of both of them, then the $K\times K$ matrix $(\mathbf{C}\boldsymbol{\Psi} \mathbf{E}_K)$ cannot be inverted and the sampling procedure will fail, regardless of the chosen node. Note that this problem is not present in classical sampling of time-varying signals, because the eigenvalues of the Fourier Vandermonde matrix associated with $\mathbf{S}=\mathbf{A}_{dc}$ are always distinct. Condition \emph{ii)} refers to the specific node where the samples of the shifted signal are observed. It basically states that any node in the network can be used to sample the signal provided that $(\mathbf{e}_k^T\boldsymbol{\upsilon}_i)\neq 0$ for $k=1, \ldots , K$; i.e., that the chosen node participates in the specific frequencies on which signal $\mathbf{x}$ is expressed. It also points to the fact that if $|\mathbf{e}_k^T\boldsymbol{\upsilon}_i|$ are non-zero but small, selecting $i$ as the sampling node may give rise to interpolations that are potentially unstable if noise is present; see Section~\ref{S:Noisy_samp_recov}. For the particular case when $\mathbf{S}=\mathbf{A}_{dc}$, condition \emph{ii)} is satisfied since all the entries of the Fourier basis are non-zero.

\subsection{Discussion}\label{Ss:DiscussionIntuitionShiftSampling}

Suppose that we know that $\mathbf{x}$ is indeed $K$-bandlimited; i.e., that it can be expressed as a linear combination of the $K$ first frequency basis vectors $\mathbf{v}_1,\ldots,\mathbf{v}_K$. Then, Proposition~\ref{P:recoveringcond_oursampling} states that a single node, say the $i$-th one, can reconstruct the entire graph signal just from its own signal $x_i$ and $K-1$ exchanges with its neighbors. Note that one of the consequences of this result is that linear combinations of signals at nodes that are in a neighborhood of radius $K-1$ suffice to reconstruct the entire graph signal. To be specific, suppose that $\mathbf{x}=\alpha\mathbf{v}_1$, which represents the extreme case of a 1-bandlimited signal. Then, it follows that $[\mathbf{y}_i]_1 = x_i = \alpha [\boldsymbol{\upsilon}_i]_1$, from where $\alpha$ can be found -- and $\mathbf{x}$ reconstructed -- as long as $[\boldsymbol{\upsilon}_i]_1 \neq 0$ [cf. condition \emph{ii)} in Proposition~\ref{P:recoveringcond_oursampling}]. This implies that a node can reconstruct a 1-bandlimited signal based solely on the value that this signal takes at the node. For the case of a 2-bandlimited signal where $\mathbf{x}=\alpha_1 \mathbf{v}_1 + \alpha_2 \mathbf{v}_2$, Proposition~\ref{P:recoveringcond_oursampling} guarantees reconstruction based on $[\mathbf{y}_i]_1$ and $[\mathbf{y}_i]_2$, which only contain information about the signal at node $i$ and at its neighbors. Therefore, one can understand bandlimited graph signals as signals that can be identified \emph{locally} by relying on observations within a given number of hops. Note that this does not necessarily imply that the variation of the signal among close-by nodes is small, it only means that the pattern of variation can be inferred just by looking at close-by nodes. This discussion will be revisited in Section~\ref{S:unknown_support}. For the \emph{recovery} to be implemented locally too, the nodes need to know $\mathbf{V}_K$ and $\{\lambda_k\}_{k=1}^K$, i.e. the structure of the graph where the signal resides.

We may decompose the interpolator $\mathbf{V}_K(\mathbf{C}\boldsymbol{\Psi}_i)^{-1}$ in \eqref{E:interp_oursampling} into three factors $\mathbf{V}_K(\mathbf{E}_K^T\diag(\boldsymbol{\upsilon}_i)\mathbf{E}_K)^{-1}(\mathbf{C}\boldsymbol{\Psi} \mathbf{E}_K)^{-1}$ to reveal that it can be computed in closed-form, since a non-zero diagonal matrix can be trivially inverted and closed-form expressions for the inverse of a Vandermonde matrix exist \cite{macon1958inverses}. Moreover, notice that one of these three factors is related to the structure of the graph and the support where the signal is bandlimited $\mathbf{V}_K$; one is related to the structure of the graph, the support of the signal and the subset of observations $(\mathbf{C}\boldsymbol{\Psi} \mathbf{E}_K)^{-1}$; and the third one depends on the specific node where the samples are taken $(\mathbf{E}_K^T\diag(\boldsymbol{\upsilon}_i)\mathbf{E}_K)^{-1}$.

\begin{remark}\label{rem_structure_observation_matrix}
\normalfont The structure of the selection matrix $\mathbf{C}$ and, in particular, the fact that $\mathbf{C}\boldsymbol{\Psi} \mathbf{E}_K$ is a Vandermonde matrix are instrumental to guarantee the recovery of the original signal. Note that $\mathbf{C}\boldsymbol{\Psi} \mathbf{E}_K$ is Vandermonde not only when $\mathbf{C}=\mathbf{E}_K^T$, but also when $\mathbf{C}=[\mathbf{e}_1,\mathbf{e}_{1+N_0},\ldots,\mathbf{e}_{1+(K-1)N_0}]^T$, provided that $1\leq N_0\leq N/K$ and $\lambda_{k_1}^{N_0}\neq\lambda_{k_2}^{N_0}$ for all $k_1\neq k_2$, where $k_1 \leq K$ and $k_2 \leq K$. By setting $N_0=N/K$, the counterpart of the classical time sampling theorem (which considers uniformly spaced samples) is recovered. Moreover, if none of the frequencies of interest is zero (i.e., if $\lambda_k\neq 0$ for $k\leq K$), then selection patterns of the form $\mathbf{C}=[\mathbf{e}_{n_0},\mathbf{e}_{n_0+N_0},\ldots,\mathbf{e}_{n_0+(K-1)N_0}]^T$ are also guaranteed to lead to invertible matrices. In this case, the resultant matrix is not Vandermonde, but it can be expressed as a product of a Vandermonde and a non-zero diagonal matrix. For reference in the following sections, we define here the $K\times N$ matrix $\mathbf{C}_K(n_0,N_0):=[\mathbf{e}_{n_0},\mathbf{e}_{n_0+N_0},\ldots,\mathbf{e}_{n_0+(K-1)N_0}]^T$ and the \emph{set} of admissible $K\times N$ selection matrices $\mathcal{C}_K:=\big\{\mathbf{C}_K(n_0,N_0)\;\;|\;N_0=1,\ldots, N/K \;\mathrm{and} \; n_0=1,\ldots, N - N_0(K -1)\big\}$. 

\end{remark}

\section{Sampling and interpolation in the presence of noise}\label{S:Noisy_samp_recov}

When sampling in the absence of noise, two main questions are how to recover the signal from its samples and the conditions under which recovery is feasible. When the samples are noisy, perfect reconstruction is, in general, unfeasible and new issues arise.
In Section~\ref{sub_sec_mvu_interpolation} we estimate the noisy signal through interpolation via the Best Linear Unbiased Estimator (BLUE) for a general noise model. In Section~\ref{subsec_error_performance_noise_models}, we specify noise models that are likely to arise in graph domains. Then, in Section~\ref{Ss:optimizing_sampling_set}, we discuss the effect on the interpolation error of selecting the sampling node and the rows of the selection matrix.

\subsection{BLUE interpolation}\label{sub_sec_mvu_interpolation}

Consider now that the shifted sampled signal $\mathbf{y}_i$ is corrupted by additive noise, so that the \emph{observed} signal $\mathbf{z}_i$ is given by $\mathbf{z}_i=\mathbf{y}_i+\mathbf{w}_i$. The noise $\mathbf{w}_i$ is assumed to be zero-mean, independent of the graph signal, and colored with a covariance matrix $\mathbf{R}_{w}^{(i)}:=\EE[\mathbf{w}_i\mathbf{w}_i^H]$. For notational convenience, we define also $\bar{\mathbf{w}}_i=\mathbf{C}\mathbf{w}_i$ and $\bar{\mathbf{R}}_w^{(i)}=\mathbf{C}\mathbf{R}_w^{(i)}\mathbf{C}^H$.

Key to design the interpolator in the presence of noise is to notice that the relation between the observed samples $\bar{\mathbf{z}}_i$ and the original signal $\mathbf{x}$ is given by
\begin{eqnarray}\label{E:linear_obs_model_noise}
\bar{\mathbf{z}}_i &=& \mathbf{C} \boldsymbol{\Psi}_i \widehat{\mathbf{x}}_K+ \bar{\mathbf{w}}_i,\label{E:linear_obs_model_noise.subeq_b}\\
\mathbf{x} &=& \mathbf{V}_K \widehat{\mathbf{x}}_K. \label{E:linear_obs_model_noise.subeq_a}
\end{eqnarray}
The BLUE estimator of $\widehat{\mathbf{x}}_K$, which minimizes the least squares error, is given by \cite{KayBook}
\begin{equation}\label{E:est_freq_noise}
\hat{\widehat{\mathbf{x}}}_K^{(i)}= \Big( \boldsymbol{\Psi}_i^H\mathbf{C}^H(\bar{\mathbf{R}}_w^{(i)})^{-1}
\mathbf{C}\boldsymbol{\Psi}_i\Big)^{-1}\boldsymbol{\Psi}_i^H\mathbf{C}^H(\bar{\mathbf{R}}_w^{(i)})^{-1}\bar{\mathbf{z}}_i,
\end{equation}
provided that the inverse in \eqref{E:est_freq_noise} exists. Additionally, for the particular case of Gaussian noise in \eqref{E:linear_obs_model_noise.subeq_b}, the estimator in \eqref{E:est_freq_noise} coincides with the Minimum Variance Unbiased (MVU) estimator which attains the Cram\'er-Rao lower bound. In this case, it also holds true that the inverse of the error covariance matrix associated with \eqref{E:est_freq_noise} corresponds to the Fisher Information Matrix (FIM) \cite{KayBook}. Clearly the larger the number of rows in \eqref{E:linear_obs_model_noise.subeq_b}, the better the estimation is. When the selection matrix $\mathbf{C}$ selects exactly $K$ rows (and not more), \eqref{E:est_freq_noise} reduces to
\begin{equation}\label{E:est_freq_noise_reduced}
\hat{\widehat{\mathbf{x}}}_K^{(i)}= \left(\mathbf{C}\boldsymbol{\Psi}_i\right)^{-1} \bar{\mathbf{z}}_i.
\end{equation}
After obtaining $\hat{\widehat{\mathbf{x}}}_K^{(i)}$ -- either via \eqref{E:est_freq_noise} or \eqref{E:est_freq_noise_reduced} --, the time signal recovered at the $i$-th node $\hat{\mathbf{x}}^{(i)}$ can be found as
\begin{equation}\label{E:est_time_noise}
\hat{\mathbf{x}}^{(i)}=\mathbf{V}_K \hat{\widehat{\mathbf{x}}}_K^{(i)}.
\end{equation}
Finally, the error covariance matrices for the frequency and time estimators $\widehat{\mathbf{R}}_e^{(i)}:=\EE[(\widehat{\mathbf{x}}_K-\hat{\widehat{\mathbf{x}}}_K^{(i)})(\widehat{\mathbf{x}}_K-\hat{\widehat{\mathbf{x}}}_K^{(i)})^H]$ and $\mathbf{R}_e^{(i)}:=\EE[(\mathbf{x}-\hat{\mathbf{x}}^{(i)})(\mathbf{x}-\hat{\mathbf{x}}^{(i)})^H]$ are \cite{KayBook}
\begin{eqnarray}
\widehat{\mathbf{R}}_e^{(i)}&=&\big( \boldsymbol{\Psi}_i^H\mathbf{C}^H(\bar{\mathbf{R}}_w^{(i)})^{-1}\mathbf{C}\boldsymbol{\Psi}_i\big)^{-1},\label{E:cov_error_est_freq_noise}\\
\mathbf{R}_e^{(i)}&=&\mathbf{V}_K \widehat{\mathbf{R}}_e^{(i)} \mathbf{V}_K^H.\label{E:cov_error_est_time_noise}
\end{eqnarray}

Note that the error covariance matrix $\mathbf{R}_e^{(i)}$ depends on the noise model, the frequencies of the graph (eigenvalues of the shift operator), the node taking the observations, and the sample-selection scheme adopted (cf. Remark~\ref{rem_structure_observation_matrix}).

The error covariance matrix enables us to assess the performance of the estimation. Smaller errors give rise to better estimators. However, there exist multiple alternatives to quantify the error, as analyzed by the theory of optimal design of experiments \cite{pukelsheim1993optimal}. The most common approach is to find an estimator which minimizes the trace of the covariance matrix
\begin{equation}\label{E:error_metric_trace}
e_1:=\mathrm{trace}(\mathbf{R}_e^{(i)}),
\end{equation}
which corresponds to the minimization of the Mean Square Error (MSE). Other common error metrics based on the error covariance matrix are the largest eigenvalue
\begin{equation}\label{E:error_metric_eigenvalue}
e_2 \!:=\!\lambda_{\max}(\mathbf{R}_e^{(i)}),
\end{equation}
the log determinant
 \begin{equation}\label{E:error_metric_logdet}
e_3 \!:= \! \log \det(\widehat{\mathbf{R}}_e^{(i)}),
\end{equation}
and the inverse of the trace of its inverse
\begin{equation}\label{E:error_metric_inverse_trace}
e_4 \!:=\! \left[\mathrm{trace}\left(\widehat{\mathbf{R}}_e^{(i)^{-1}}\right)\right]^{-1}.
\end{equation}
Notice that the error metrics $e_3$ and $e_4$ are computed based on the error covariance matrix for the frequency estimator $\widehat{\mathbf{R}}_e^{(i)}$ instead of the time estimator since $\mathbf{R}_e^{(i)}$ is a singular matrix [cf. \eqref{E:cov_error_est_time_noise}].


\subsection{Noise models}\label{subsec_error_performance_noise_models}

The results presented so far consider a general error covariance matrix $\mathbf{R}_w^{(i)}$, so that they can be used regardless of the color of the noise. In this section, we present three particular examples of interest.
\begin{itemize}
\item White noise in the observed signal $\mathbf{z}_i$. This implies that $\mathbf{w}_i$ is white and therefore $\mathbf{R}_w^{(i)} = \sigma^2 \mathbf{I}$, with $\sigma^2$ denoting the noise power. In this case, the $K \!\times\! K$ matrix $\bar{\mathbf{R}}_w^{(i)}$ is given by
\begin{equation}\label{E:noise_model_1}
\bar{\mathbf{R}}_w^{(i)} = \sigma^2 \mathbf{I}.
\end{equation}
\item White noise in the original signal $\mathbf{x}$. With $\mathbf{w}$ denoting the white additive noise present in $\mathbf{x}$, we can use the linear observation model to write $\mathbf{w}_i=\boldsymbol{\Psi} \diag(\boldsymbol{\upsilon}_i)\mathbf{V}^{-1}\mathbf{w}$. Then, the $N\times N$ error correlation matrix is simply given by $\mathbf{R}_w^{(i)} = \sigma^2 \boldsymbol{\Psi} \diag(\boldsymbol{\upsilon}_i)\mathbf{V}^{-1} (\mathbf{V}^{-1})^H  \diag(\boldsymbol{\upsilon}_i)^H \boldsymbol{\Psi}^H$. When the shift is a normal matrix, $\mathbf{V}$ is unitary and the previous expression reduces to $\mathbf{R}_w^{(i)} = \sigma^2 \boldsymbol{\Psi} |\diag(\boldsymbol{\upsilon}_i)|^2 \boldsymbol{\Psi}^H$. As before, the $K \times K$ error correlation matrix is obtained just by selecting the rows and columns of the former,
\begin{equation}\label{E:noise_model_2}
\bar{\mathbf{R}}_w^{(i)} = \sigma^2  \mathbf{C} \boldsymbol{\Psi} |\diag(\boldsymbol{\upsilon}_i)|^2 \boldsymbol{\Psi}^H\mathbf{C}^H.
\end{equation}
The previous expressions show not only that the noise is correlated, but also that the correlation depends on the graph structure (eigenvalues and eigenvectors of $\mathbf{S}$), the node collecting the observations, and the specific selection of observations.
\item White noise in the active frequency coefficients $\widehat{\mathbf{x}}_K$. With $\widehat{\mathbf{w}}_K$ denoting the white additive noise present in $\widehat{\mathbf{x}}_K$, we can use the linear observation model to write $\mathbf{w}_i=\boldsymbol{\Psi} \diag(\boldsymbol{\upsilon}_i)\mathbf{E}_K\widehat{\mathbf{w}}_K\!=\!\boldsymbol{\Psi}_i\widehat{\mathbf{w}}_K$. It follows that the $N\! \times \!N$ and $K \!\times \!K$ error covariance matrices are $\mathbf{R}_w^{(i)} = \sigma^2 \boldsymbol{\Psi}_i \boldsymbol{\Psi}_i^H$ and
\begin{equation}\label{E:noise_model_3}
\bar{\mathbf{R}}_w^{(i)} = \sigma^2  \mathbf{C} \boldsymbol{\Psi}_i \boldsymbol{\Psi}_i^H \mathbf{C}^H.
\end{equation}
This model can be appropriate for scenarios where the signal of interest is the output of a given ``graph process'' -- e.g., a diffusion process -- and the noise is present in the input of that process. This noise model can also arise when the signal to be sampled has been previously processed with a low-pass graph filter \cite{SandryMouraSPG_TSP14Freq,Eusipco15}.
\end{itemize}
There are many other noise models that can be of interest in graph setups. For example, the error covariance can be a linear combination of the previous covariance matrices (noise is present in both the original signal and the observation process). Alternatively, the noise at a specific node can be also rendered dependent on the number of neighbors. This last situation would be reasonable, for example, in distributed setups where the information of neighboring nodes is exchanged via noisy channels.

\subsection{Selection of the sampling set}\label{Ss:optimizing_sampling_set}

The two elements that define the set of samples to be interpolated are: the sampling node, i.e., the node $i$ which aggregates the information; and the sample-selection scheme, i.e., the elements of $\mathbf{y}_i$ selected by $\mathbf{C}$.

\subsubsection{Selection of the sampling node}\label{subsubsec_sampling_node}
The recovery results in Section~\ref{S:sampling_GS_local_aggregat} show that any node $i$ can be used to sample and recover the entire graph signal, provided that the entries of $\boldsymbol{\upsilon}_i$ corresponding to the active frequencies in $\widehat{\mathbf{x}}$ are non-zero. However, when noise is present, the error covariance matrix $\mathbf{R}_e^{(i)}$ -- which is the key element to evaluate the quality of the interpolation -- is different for each $i$. In this context, it is reasonable to select as a sampling node one leading to a small error. Note that selecting the best one will only require the computation of $N$ closed-form expressions which involve matrix inversions. In scenarios where computational complexity is a limiting factor, the structure of the noise correlation, as well as the structure of the interpolation matrix, can be exploited to reduce considerably the computational burden.
E.g., for the case where white noise is present in the active frequency coefficients, when substituting \eqref{E:noise_model_3} into \eqref{E:cov_error_est_freq_noise} and \eqref{E:cov_error_est_time_noise}, it follows that
\begin{equation}\label{E:node_choice_noise_3}
\widehat{\mathbf{R}}_e^{(i)}= \sigma^2 \mathbf{I}, \quad \mathbf{R}_e^{(i)} = \sigma^2 \mathbf{V}_K \mathbf{V}_K^H.
\end{equation}
Consequently, for this particular noise model, the estimator performance is independent of the node choice. This is true for every error metric [cf. \eqref{E:error_metric_trace}-\eqref{E:error_metric_inverse_trace}]. The result is intuitive: given that the noise and the signal are present in the same frequencies, it is irrelevant if a node amplifies or attenuates a particular frequency. Differently, if the white noise is present in the observed signal, we can substitute \eqref{E:noise_model_1} into \eqref{E:cov_error_est_freq_noise} to obtain
\begin{equation}\label{E:node_choice_noise_1}
\widehat{\mathbf{R}}_e^{(i)} = \sigma^2 \big( \mathbf{E}_K^H \diag(\boldsymbol{\upsilon}_i)^H \boldsymbol{\Psi}^H\mathbf{C}^H \mathbf{C} \boldsymbol{\Psi} \diag(\boldsymbol{\upsilon}_i)\mathbf{E}_K \big)^{-1}.
\end{equation}
Thus, if we are interested in minimizing, e.g., the error metric $e_4$ [cf. \eqref{E:error_metric_inverse_trace}], our objective may be reformulated as finding the optimal node $i^*$ such that
\begin{equation}\label{E:optimal_node_choice_noise_1}
i^* \! = \! \arg\max_{i} \,\, \mathrm{trace} \left( \mathbf{E}_K^H \diag(\boldsymbol{\upsilon}_i)^H \boldsymbol{\Psi}^H\mathbf{C}^H \mathbf{C} \boldsymbol{\Psi} \diag(\boldsymbol{\upsilon}_i)\mathbf{E}_K \right).
\end{equation}
For a selection matrix of the form $\mathbf{C} = \mathbf{C}_{K}(n_0, N_0)$ (cf. Remark~\ref{rem_structure_observation_matrix}), the $k$-th diagonal element of the matrix in \eqref{E:optimal_node_choice_noise_1} can be written as $|[\boldsymbol{\upsilon}_i]_k|^2 \sum_{m=0}^{K-1} |\lambda_k|^{2 \, (n_0 + m N_0)}$. The trace is simply the sum of those elements, so that, using the closed form for a geometric sum, \eqref{E:optimal_node_choice_noise_1} can be rewritten as
%
\begin{equation}\label{E:optimal_node_choice_noise_2}
i^* \! = \! \arg\max_{i} \,\,\, \sum_{k=1}^K \, |[\boldsymbol{\upsilon}_i]_k|^2 \frac{|\lambda_k|^{2n_0}-|\lambda_k|^{2(n_0+N_0K)}}{1-|\lambda_k|^{2N_0}}.
\end{equation}
Thus, the optimal sampling node $i^*$ will be one with large values of $|[\boldsymbol{\upsilon}_i]_k|$ for the active frequencies $k \leq K$. The relative importance of frequency $k$ is given by the fraction in \eqref{E:optimal_node_choice_noise_2}, which depends on the modulus of the associated eigenvalue and the structure of the selection matrix $\mathbf{C}$ (values of $n_0$ and $N_0$).

\subsubsection{Design of the sample-selection scheme}\label{subsubsec_sampled_shifts}
The error covariance matrix, and hence the different error metrics presented in \eqref{E:error_metric_trace}-\eqref{E:error_metric_inverse_trace}, depend on the selection matrix $\mathbf{C}$. By changing $\mathbf{C}$ one can tradeoff the quality of a given sample and the detrimental effect of the corresponding noise. The specific set of samples that minimizes the error will in general depend on the error metric chosen.


Recall that any matrix in the set of admissible selection matrices $\mathcal{C}_K$ defined in Remark~\ref{rem_structure_observation_matrix} is guaranteed to lead to a feasible recovery according to the conditions stated in Proposition~\ref{P:recoveringcond_oursampling}. If $\mathbf{C}$ is not constrained to belong to $\mathcal{C}_K$, the number of candidate matrices is $N$ choose $K$. However, $\mathcal{C}_K$ has a much smaller cardinality: $N_0$ can take at most $N/K$ values, and $n_0$ at most $(N - N_0(K -1))$.
Moreover, as it was the case for the sampling node selection, in some cases the noise structure can be exploited to readily determine the optimal observation strategy. E.g., for the case where white noise is present in the active frequencies, it is immediate to see that the performance is independent of the sample-selection scheme [cf. \eqref{E:node_choice_noise_3}]. For the case where white noise is present in the observed signal, let us assume that the selection matrix is given by $\mathbf{C} = \mathbf{C}_K(n_0, N_0)$ where $N_0$ is fixed and we want to design $n_0$.

If we adopt $e_3$ in \eqref{E:error_metric_logdet} as our error metric, the goal is to find the value $n_0^*$ that minimizes $\mathrm{det}\big( \widehat{\mathbf{R}}_e^{(i)}\big)$. To achieve this, consider two different selection matrices $\mathbf{C}_A = \mathbf{C}_K(n_0, N_0)$ and $\mathbf{C}_B = \mathbf{C}_K(n_0+1, N_0)$. Using \eqref{E:node_choice_noise_1} and assuming without loss of generality that $\sigma^2=1$, the error covariance for $\mathbf{C}_B$ is given by $\widehat{\mathbf{R}}_{e,B}^{(i)}=(\mathbf{E}_K^H\diag(\boldsymbol{\upsilon}_i)^H\boldsymbol{\Psi}^H\mathbf{C}_B^H\mathbf{C}_B \boldsymbol{\Psi}\diag(\boldsymbol{\upsilon}_i)\mathbf{E}_K)^{-1}$. A similar expression can be written for $\widehat{\mathbf{R}}_{e,A}^{(i)}$. Since $\boldsymbol{\Psi}$ is Vandermonde, it is not difficult to show that $\boldsymbol{\Psi}^H\mathbf{C}_B^H\mathbf{C}_B \boldsymbol{\Psi}$ can be written as $\boldsymbol{\Lambda}^H\boldsymbol{\Psi}^H\mathbf{C}_A^H\mathbf{C}_A \boldsymbol{\Psi}\boldsymbol{\Lambda}$. This implies that
\begin{align}
\widehat{\mathbf{R}}_{e,B}^{(i)^{-1}}&= \mathbf{E}_K^H \boldsymbol{\Lambda}^H \diag(\boldsymbol{\upsilon}_i)^H\boldsymbol{\Psi}^H\mathbf{C}_A^H\mathbf{C}_A \boldsymbol{\Psi}\diag(\boldsymbol{\upsilon}_i)\boldsymbol{\Lambda} \mathbf{E}_K \nonumber \\
&=(\mathbf{E}_K^H \boldsymbol{\Lambda}^H \mathbf{E}_K) \widehat{\mathbf{R}}_{e,A}^{(i)^{-1}} (\mathbf{E}_K^H \boldsymbol{\Lambda} \mathbf{E}_K).\label{E:relation_error_cov_CA_CB}
\end{align}
For the first equality we have used that the product of diagonal matrices is commutative and for the second one that right and left multiplying by the canonical matrix amounts to selecting the columns and rows of the multiplied matrix. Using \eqref{E:relation_error_cov_CA_CB}, we have that
\begin{equation}\label{E:relation_determinants_CA_CB}
\mathrm{det} \big(\widehat{\mathbf{R}}_{e,A}^{(i)}\big) =  \mathrm{det}\!\big( \widehat{\mathbf{R}}_{e,B}^{(i)} \big) \prod_{k=1}^K  |\lambda_k|^2,
\end{equation}
which results in the following optimal strategy for the solution of $e_3$: if $\prod_{k=1}^K |\lambda_k|^2 \leq 1$ then $n_0^* = 1$, otherwise $n_0^*$ should be as large as possible; see Remark~\ref{rem_length_observation_matrix}. Equivalently, the optimal strategy states that if an application of the shift operator has an overall effect of amplification in the active frequencies, then we should aim to apply it as many times as possible, whereas if the opposite is true, we should avoid its application.

Needless to say, one can also look at selection matrices that are not always guaranteed to lead to a Vandermonde structure, i.e., matrices not in $\mathcal{C}_K$. In that case, the problem can be formulated as a binary optimization over $\mathbf{C}$, which is typically challenging. If the size of the space search ($N$ choose $K$) is not too large, the problem can be solved by exhaustive search -- first by checking that the matrix guarantees recovery and then evaluating the corresponding error covariance. For more general cases, a reasonable approach is to formulate the problem, relax it, and exploit the problem structure to find a good approximate solution efficiently. The problem formulation and the subsequent relaxation will depend on the specific optimality criteria. Although of interest, developing approximate algorithms to design the selection matrix $\mathbf{C}$ is out of the scope of this paper and is left as future work.

%
%

It is worth stressing that the sample-selection scheme that minimizes the error does not have to be the same for all nodes. Both the selection of the sampling node and the sampling shifts can be combined to obtain the best local reconstruction across all nodes in the graph.

\begin{remark}\label{rem_length_observation_matrix}
Designing $\mathbf{C}$ entails the selection of a subset of $K$ entries out of the $N$ entries in $\mathbf{y}_i$. However, $\mathbf{y}_i$ has only $N$ entries because $\mathbf{Y}$ has only $N$ columns [cf. \eqref{eqn_all_shifts_matrix}]. Strictly speaking, there is no need to impose this restriction and more columns could be added to $\mathbf{Y}$. As a matter of fact, if for a given \emph{noisy} graph signal the application of the shift operator $\mathbf{S}$ attenuates the noise while amplifying the signal, the sampling procedure will benefit from further applications of $\mathbf{S}$, even beyond the size of the graph $N$. In practice, the maximum number of applications will be limited by the computational and signaling cost associated with the application of the shift.
\end{remark}

\section{Identifying the support of the graph signal}\label{S:unknown_support}

In the previous sections, it has been assumed that the frequency support of the bandlimited signal corresponded to the $K$ principal eigenvectors, which are the ones associated with the largest eigenvalues. However, the results presented also hold true as long as the basis support, i.e., the frequencies that are present in $\mathbf{x}$, are known. To be specific, let $\mathcal{K}:=\{k_1,\ldots,k_K\}$ denote the set of indexes where the signal $\mathbf{x}$ is sparse and, based on it, define the $N\times K$ matrices $\mathbf{V}_{\mathcal{K}}:=[\mathbf{v}_{k_1},\ldots,\mathbf{v}_{k_K}]$ and $\mathbf{E}_{\mathcal{K}}:=[\mathbf{e}_{k_1},\ldots,\mathbf{e}_{k_K}]$. Then, all the results presented so far hold true if $\mathbf{V}_K$ is replaced with $\mathbf{V}_{\mathcal{K}}$, and $\mathbf{E}_K$, when used to select the active frequencies, is replaced with $\mathbf{E}_{\mathcal{K}}$.

A related but more challenging problem is to design the sampling and interpolation procedures when the frequency support $\mathcal{K}$ is not known. Generically, this problem falls into the class of sparse signal reconstruction \cite{donoho2003spark,elad2007optimized,RabICASSP12_SpectCompressSensingGraphs}. However, the particularities of our setup can be exploited to achieve stronger results. In particular, for the sampling procedure proposed in this paper, the so-called sensing matrix -- the one relating the signal of interest to the observed samples -- has a useful Vandermonde structure that can be exploited.

\subsection{Noiseless joint recovery and support identification}\label{Ss:noiseless_joint_recovery_support_identification}
Consider the noiseless aggregation sampling of Section~\ref{S:sampling_GS_local_aggregat}, where we know that $\widehat{\mathbf{x}}$ is $K$-sparse but we do not know the support of the $K$ non-zero entries [cf. \eqref{E:sampled_shifted_signal}]
\begin{equation}\label{E:sampled_shifted_signal_unkown_v0}
\bar{\mathbf{y}}_i=\mathbf{C}\mathbf{y}_i=\mathbf{C}\boldsymbol{\Psi}\diag(\boldsymbol{\upsilon}_i)\widehat{\mathbf{x}}.
\end{equation}
For the case where the support is known, it was shown that a selection matrix $\mathbf{C}$ that picks the first $K$ rows of $\boldsymbol{\Psi}$ is enough for perfect reconstruction (cf. Proposition~\ref{P:recoveringcond_oursampling}).

If we reformulate the recovery problem as
\begin{align}\label{E:sampled_shifted_signal_unkown}
\widehat{\mathbf{x}}^*:=&\arg\min_{\widehat{\mathbf{x}}} \quad \quad ||\widehat{\mathbf{x}}||_0 \\
&\mathrm{s. t.} \quad\quad \mathbf{C}\mathbf{y}_i= \mathbf{C}\boldsymbol{\Psi}\diag(\boldsymbol{\upsilon}_i)\widehat{\mathbf{x}}, \nonumber
\end{align}
for the unknown support case, there is no guarantee that the solution $\widehat{\mathbf{x}}^*$ coincides with the $K$-sparse representation of the observed signal. When the frequency support is known, and provided that the selection matrix satisfies the conditions in Remark~\ref{rem_structure_observation_matrix}, selecting $K$ rows of $\boldsymbol{\Psi}\diag(\boldsymbol{\upsilon}_i)\mathbf{E}_{\mathcal{K}}$ leads to a (one-to-one) invertible transformation. When the support is unknown, guaranteeing identifiability requires selecting a higher number of rows (samples) \cite{donoho2003spark}. The following proposition states this result formally. To simplify notation, the proposition assumes that $K\leq N/2$, but the result holds true also when that is not the case.
\begin{myproposition}\label{P:joint_recovery_supportid}
\emph{Let $\mathbf{x}$ and $\mathbf{C}$ be, respectively, a bandlimited graph signal with at most $K$ non-zero frequency components and a selection matrix with $2K$ rows of the form $\mathbf{C}=\mathbf{C}_{2K}(n_0,N_0)$ (cf. Remark~\ref{rem_structure_observation_matrix}). Then, if all the entries in $\boldsymbol{\upsilon}_i$ are non-zero and all the eigenvalues of $\mathbf{S}$ are non-zero and satisfy that $\lambda_{k}^{N_0}\neq\lambda_{k'}^{N_0}$ for all $k\neq k'$, it holds that\\
\noindent i) the solution to \eqref{E:sampled_shifted_signal_unkown} is unique; and\\
\noindent ii) the original graph signal can be recovered as $\mathbf{x}=\mathbf{V}\widehat{\mathbf{x}}^*$.}
\end{myproposition}
\begin{myproof}
The proof proceeds into two steps. The first step is to show that any selection of $2K$ columns of the $2K \times N$ matrix $\mathbf{M}:=\mathbf{C}\boldsymbol{\Psi}\diag(\boldsymbol{\upsilon}_i)$ has rank $2K$ and, hence, it leads to an invertible $2K \times 2K$ matrix. To prove this, let $\mathcal{F}=\{f_1,\ldots,f_{2K}\}$ be a set with cardinality $2K$ containing the indexes of the selected columns and define the $N\times 2K$ canonical matrix $\mathbf{E}_{\mathcal{F}}=[\mathbf{e}_{f_1},\ldots,\mathbf{e}_{f_{2K}}]$. Using this notation, the matrix containing the columns of $\mathbf{M}$ indexed by $\mathcal{F}$ is $\mathbf{M}\mathbf{E}_{\mathcal{F}}$, which can be alternatively written as
\begin{equation}
\mathbf{M}\mathbf{E}_{\mathcal{F}}=\mathbf{C}\boldsymbol{\Psi}\diag(\boldsymbol{\upsilon}_i)\mathbf{E}_{\mathcal{F}}=(\mathbf{C}\boldsymbol{\Psi}\mathbf{E}_{\mathcal{F}})(\mathbf{E}_{\mathcal{F}}^T\diag(\boldsymbol{\upsilon}_i)\mathbf{E}_{\mathcal{F}}).
\end{equation}
The expression reveals that $\mathbf{M}\mathbf{E}_{\mathcal{F}}$ is invertible because it can be written as the product of two $2K\times 2K$ invertible matrices. The latter is true because: a) conditions $\mathbf{C}=\mathbf{C}_{2K}(n_0,N_0)$, $\lambda_k^{N_0}\neq\lambda_{k'}^{N_0}$ for all $k\neq k'$, and $\lambda_k\neq 0$ for all $k$ guarantee that $(\mathbf{C}\boldsymbol{\Psi}\mathbf{E}_{\mathcal{F}})$ is invertible because it is a product of a diagonal and a full rank Vandermonde matrix (cf. Remark \ref{rem_structure_observation_matrix}) and b) condition $[\boldsymbol{\upsilon}_i]_k \neq 0$ for all $k$ guarantees that $(\mathbf{E}_{\mathcal{F}}^T\diag(\boldsymbol{\upsilon}_i)\mathbf{E}_{\mathcal{F}})$ is an invertible diagonal matrix. This is true for any $\mathcal{F}$. The second step is to show that $2K$ observations are enough to guarantee identifiability. To see why this is the case, assume that two different feasible solutions $\widehat{\mathbf{x}}_A$ and $\widehat{\mathbf{x}}_B$ exist. This would imply that $\mathbf{M} ( \widehat{\mathbf{x}}_A - \widehat{\mathbf{x}}_B) = 0$. Nevertheless, the vector $( \widehat{\mathbf{x}}_A - \widehat{\mathbf{x}}_B) $ has, at most, $2K$ non-zero components and any choice of $2K$ columns of $\mathbf{M}$ generates a full rank square matrix which forces $\widehat{\mathbf{x}}_A = \widehat{\mathbf{x}}_B$, contradicting the assumption of multiple solutions. Finally, it is worth noting that although the proposition requires all the eigenvalues to be non-zero and distinct, only the ones associated with $\mathcal{K}$ need to satisfy those requirements. Note that the previous proof amounts to say that the matrix $\mathbf{C}\boldsymbol{\Psi}\diag(\boldsymbol{\upsilon}_i)$ has full spark and, hence, the claims in the proposition coincide with those in \cite{donoho2003spark} for the $0$-norm recovery.
\end{myproof}

It is worth stressing that the proof for the joint recovery and identification support leverages once more the fact that $\boldsymbol{\Psi}$ is a Vandermonde matrix, which is a distinct feature of the aggregation sampling scheme proposed in this paper. To gain more intuition about the result, we revisit the discussion provided after Proposition~\ref{P:recoveringcond_oursampling} and suppose that we know that $\mathbf{x}$ is a bandlimited signal with \emph{only one} non-zero frequency component. This means that $K=1$ and that the graph signal can be written as $\mathbf{x}=\alpha \mathbf{v}_k$. If the value of $k$ is known, which amounts to say that the support where the signal is sparse is known, then node $i$ can interpolate the entire signal $\mathbf{x}$ using $x_i$ (cf. Section~\ref{Ss:DiscussionIntuitionShiftSampling}). If the support is not known, Proposition~\ref{P:joint_recovery_supportid} establishes $2K=2$ samples are needed. Clearly, one sample is not enough because $x_i=\hat{\alpha}[\mathbf{v}_k]_i$ admits $N$ different solutions, one per $k$. To see why two samples suffice, note that the first shift corresponds to $[\mathbf{y}_i]_1=x_i$ and the second to $[\mathbf{y}_i]_2=S_{ii}x_i+\sum_{j\in \mathcal{N}_i} S_{ij} x_{j}=\lambda_{\hat{k}} x_i$. Then, node $i$ can identify first the active frequency by finding the frequency index $\hat{k}$ whose associated eigenvalue is $[\mathbf{y}_i]_2/[\mathbf{y}_i]_1$. For the identification to succeed, the eigenvalues need to be distinct, as required by Proposition~\ref{P:joint_recovery_supportid}. Once the active frequency is known, the corresponding frequency coefficient can be estimated as before by setting $\hat{\alpha}=x_i/[\mathbf{v}_{\hat{k}}]_i$ and then the entire graph signal is $\hat{\mathbf{x}}=\hat{\alpha} \mathbf{v}_{\hat{k}}$. This discussion provides additional support to the idea that bandlimited graph signals can be understood as signals that can be inferred from local information.

From a computational perspective, the presence of the $0$-norm in \eqref{E:sampled_shifted_signal_unkown} renders the optimization non-convex, thus challenging to solve. A straightforward way to convexify it is to replace the $0$-norm with a $1$-norm. Note that if such a process finds a feasible solution, call it $\widehat{\mathbf{x}}_1^*$, such that $||\widehat{\mathbf{x}}_1^*||_0 = K$, then it holds that $\widehat{\mathbf{x}}^*=\widehat{\mathbf{x}}_1^*$. Conditions under which this process is guaranteed to identify the support can be found by analyzing the coherence and the restricted isometry property (RIP) of the matrix $\mathbf{C}\boldsymbol{\Psi}\diag(\boldsymbol{\upsilon}_i)$ \cite{donoho2003spark,candes2006robust}. Unfortunately, determining the conditioning of all submatrices of a deterministic matrix (and, hence, the RIP) is challenging \cite{fullsparkframes12}. The coherence of the matrix $\mathbf{C}\boldsymbol{\Psi}\diag(\boldsymbol{\upsilon}_i)$, denoted as $\mu_i(\mathbf{C})$, is easier to find and it depends on the most similar pair of eigenvalues of $\mathbf{S}$. However, the sparsity bound given by the matrix coherence, which requires $K\leq \frac{1}{2}\big(1+\frac{1}{\mu_i(\mathbf{C})}\big)$ \cite{donoho2003spark,elad2007optimized}, is oftentimes too restrictive. A better alternative in that case is to use the concept of $t$-averaged mutual coherence and apply the results in \cite{elad2007optimized} for deterministic sensing matrices.

\subsection{Noisy joint recovery and support identification}
If noise is present and the frequency support of the signal is unknown, the ($K$-sparse) least squares estimate of $\widehat{\mathbf{\mathbf{x}}}$ can be found as the solution to the following optimization problem
%
\begin{align}\label{E:sampled_shifted_signal_unkown_noise}
\widehat{\mathbf{x}}^*:=&\arg\min_{\widehat{\mathbf{x}}} \|(\bar{\mathbf{R}}_w^{(i)})^{-1/2}\big(\mathbf{C}\mathbf{y}_i - \mathbf{C}\boldsymbol{\Psi}\diag(\boldsymbol{\upsilon}_i)\widehat{\mathbf{x}}\big) \|_{2}^2\\
&\mathrm{s. t.} \quad\quad ||\widehat{\mathbf{x}}||_0 \leq K \nonumber
\end{align}
where the matrix multiplication $(\bar{\mathbf{R}}_w^{(i)})^{-1/2}$ in the objective accounts for the fact of the noise being colored. As in the noiseless case, a straightforward approach to convexify the problem is to replace the $0$-norm with the $1$-norm and solve the problem $\widehat{\mathbf{x}}_1^*:=\arg \min_{\widehat{\mathbf{x}}}  \|(\bar{\mathbf{R}}_w^{(i)})^{-1/2}\big(\mathbf{C}\mathbf{y}_i - \mathbf{C}\boldsymbol{\Psi}\diag(\boldsymbol{\upsilon}_i)\widehat{\mathbf{x}}\big) \|_2^2 + \gamma ||\widehat{\mathbf{x}}||_1$ for different values of the parameter $\gamma$.

The challenges for support identification and the penalty paid in terms of error performance are related to those in the previous sections \cite{donoho2003spark,candes2006robust,elad2007optimized}. If the conditioning of matrix $\mathbf{C}\boldsymbol{\Psi}\diag(\boldsymbol{\upsilon}_i)$ is poor, which depends heavily on how similar the eigenvalues in $\boldsymbol{\Lambda}$ are, the performance will be bad. Bounds can be found using the coherence of $\mathbf{C}\boldsymbol{\Psi}\diag(\boldsymbol{\upsilon}_i)$, which is tractable, or by analyzing its RIP. The results in \cite{elad2007optimized} for deterministic matrices can also be used here. An alternative to have performance guarantees in this case is to consider that the matrix $\mathbf{C}\boldsymbol{\Psi}\diag(\boldsymbol{\upsilon}_i)$ is random. This can be the case if, for example, $\mathbf{C}$ is designed as random or if there is noise in the application of the shift operator $\mathbf{S}$.

\section{Space-shift sampling of graph signals}\label{S:shift_space_bigvec_sampling}

This section presents an alternative -- more general -- sampling setup that combines the selection sampling presented in Section~\ref{Ss:conventional sampling} with the aggregation sampling proposed in Section~\ref{S:sampling_GS_local_aggregat}.

Let us start by defining $\mathbf{Z}=\mathbf{Y}+\mathbf{W}$ as the noisy counterpart of $\mathbf{Y}$ [cf. \eqref{eqn_all_shifts_matrix}]. Clearly, $\mathbf{z}_i$ -- the observed shifted signal at node $i$ -- corresponds to a row of matrix $\mathbf{Z}$. We are now interested in collecting samples at different nodes and shifts, i.e., we want to sample matrix $\mathbf{Z}$. To do so, we first define the vectorized version of $\mathbf{Z}$ as $\underset{\bar{}}{\mathbf{z}}=\mathrm{vec}(\mathbf{Z}^T)$. Recall that signal $\mathbf{z}_i$ can be related to $\widehat{\mathbf{x}}_K$ via $\mathbf{z}_i=\boldsymbol{\Psi} \mathbf{E}_K\diag(\boldsymbol{\bar{\upsilon}}_i)\widehat{\mathbf{x}}_K+\mathbf{w}_i$ [cf. \eqref{E:linear_obs_model_noise}], where $\boldsymbol{\bar{\upsilon}}_i = \boldsymbol{\upsilon}_i\mathbf{E}_K$. To write a similar equation relating $\underset{\bar{}}{\mathbf{z}}$ to $\widehat{\mathbf{x}}_K$, we need to define the $N^2 \times N$ matrix $\boldsymbol{\Upsilon}$ and its corresponding reduced $NK \times K$ matrix $\bar{\boldsymbol{\Upsilon}}$ as
\begin{equation}\label{eqn_def_bar_upsilon}
\boldsymbol{\Upsilon}:= \left( \begin{array}{c}
\diag(\boldsymbol{\upsilon}_1) \\
\vdots  \\
\diag(\boldsymbol{\upsilon}_N) \end{array} \right), \quad\quad
\bar{\boldsymbol{\Upsilon}}:= \left( \begin{array}{c}
\diag(\bar{\boldsymbol{\upsilon}}_1) \\
\vdots  \\
\diag(\bar{\boldsymbol{\upsilon}}_N) \end{array} \right).
\end{equation}
Based on this, $\underset{\bar{}}{\mathbf{z}}$ can be written as
\begin{equation}\label{eqn_augmented_system_of_equations}
\underset{\bar{}}{\mathbf{z}}=\Big(\mathbf{I}\otimes(\boldsymbol{\Psi} \mathbf{E}_K)\Big) \bar{\boldsymbol{\Upsilon}} \widehat{\mathbf{x}}_K + \underset{\bar{}}{\mathbf{w}},
\end{equation}
where $\underset{\bar{}}{\mathbf{w}}$ is a vector of length $N^2$ obtained by concatenating the noise vectors $\mathbf{w}_i$ for all nodes $i$. This implies that \eqref{eqn_augmented_system_of_equations} is a system of $N^2$ linear equations with $K < N$ variables.
Thus, our objective is to pick $K$ of these equations in order to estimate $\widehat{\mathbf{x}}_K$ -- and, hence, $\mathbf{x}$ through \eqref{E:est_time_noise} -- while minimizing the error introduced by the noise $\underset{\bar{}}{\mathbf{w}}$. Notice that if we restrict ourselves to pick $K$ equations out of the $N$ equations in positions $(i-1)\,N+1$ to $i \, N$ for some node $i \in \{1, 2, \ldots, N\}$, then the problem reduces to local aggregation sampling at node $i$ as developed in Section~\ref{S:sampling_GS_local_aggregat}. Similarly, if we restrict ourselves to pick the $K$ equations out of the $N$ equations in positions $[1, 1+N, 1+2\,N, \ldots, 1+(N-1)\,N]$, then the problem reduces to selection sampling as presented in Section~\ref{Ss:conventional sampling}. In this sense, the formulation in \eqref{eqn_augmented_system_of_equations} is more general. To implement the selection of the $K$ equations out of the $N^2$ options in \eqref{eqn_augmented_system_of_equations}, we use a binary selection matrix $\bbC$ as done in previous sections but, in this case, the size of $\mathbf{C}$ is $K \times N^2$. The reduced square system of linear equations can then be written as [cf. \eqref{E:linear_obs_model_noise}]
\begin{equation}\label{eqn_augmented_reduced_system_of_equations}
\bar{\underset{\bar{}}{\mathbf{z}}}= \mathbf{C} \Big(\mathbf{I}\otimes(\boldsymbol{\Psi} \mathbf{E}_K)\Big) \bar{\boldsymbol{\Upsilon}} \widehat{\mathbf{x}}_K + \mathbf{C} \underset{\bar{}}{\mathbf{w}}.
\end{equation}
The correlation matrices of the frequency $\widehat{\mathbf{R}}_{\underset{\bar{}}{e}}$ and time $\mathbf{R}_{\underset{\bar{}}{e}}$ errors of the estimator computed as the solution of \eqref{eqn_augmented_reduced_system_of_equations} are [cf. \eqref{E:cov_error_est_freq_noise} and \eqref{E:cov_error_est_time_noise}]
\begin{align}
\widehat{\mathbf{R}}_{\underset{\bar{}}{e}}&=\Big(   \bar{\boldsymbol{\Upsilon}}^H  \left(\mathbf{I}\otimes(\boldsymbol{\Psi} \mathbf{E}_K)\right)^H \mathbf{C}^H     \nonumber \\
   &\times (\mathbf{C} \mathbf{R}_{\underset{\bar{}}{w}} \mathbf{C}^H)^{-1} \times \mathbf{C} \left(\mathbf{I}\otimes(\boldsymbol{\Psi} \mathbf{E}_K)\right) \bar{\boldsymbol{\Upsilon}}    \Big)^{-1}, \label{E:augmented_cov_error_est_freq_noise}\\
\mathbf{R}_{\underset{\bar{}}{e}}&=\mathbf{V}_K \widehat{\mathbf{R}}_{\underset{\bar{}}{e}} \mathbf{V}_K^H, \label{E:augmented_cov_error_est_time_noise}
\end{align}
where $\mathbf{R}_{\underset{\bar{}}{w}}$ is the covariance matrix of the stacked vector of noise $\underset{\bar{}}{\mathbf{w}}$. In this aggregated case, the same noise models introduced in Section~\ref{subsec_error_performance_noise_models} can be present. For the white noise in observations, we have that $\mathbf{R}_{\underset{\bar{}}{w}}=\sigma^2 \mathbf{I}$, for the white noise in the original signal, we have that $\mathbf{R}_{\underset{\bar{}}{w}} = \sigma^2 \left(\mathbf{I}\otimes\boldsymbol{\Psi}\right) \boldsymbol{\Upsilon} \boldsymbol{\Upsilon}^H \left(\mathbf{I}\otimes\boldsymbol{\Psi}\right)^H$, and for the white noise in the active frequency coefficients we have that $\mathbf{R}_{\underset{\bar{}}{w}} = \sigma^2 \left(\mathbf{I}\otimes (\boldsymbol{\Psi} \mathbf{E}_K)\right) \bar{\boldsymbol{\Upsilon}} \bar{\boldsymbol{\Upsilon}}^H \left(\mathbf{I}\otimes (\boldsymbol{\Psi} \mathbf{E}_K)\right)^H$.

\subsection{Structured observability pattern}

In the previous discussion, no structure was assumed in the selection matrix $\mathbf{C}$. A case of particular interest is when the sampling schemes are implemented in a distributed manner using message passing. Suppose that the sampling is performed at node $i$. To compute $y_i^{(l)}$, the node $i$ needs to have access to $y_j^{(l')}$ for all $j\in \mathcal{N}_i$ and $l'<l$. To simplify notation, and without loss of generality, we will assume that the sampling node is $i=1$ and that the neighbors of $i\!=\!1$ are $i=2,\ldots,N_1+1$. Suppose also that node $i\!=\!1$ computes $L_1$ shifts, from $y_1^{(0)}$ up to $y_1^{(L_1)}$. This implies that node $i=1$ has access to $L_1+1$ of its own samples and to $L_1$ samples of each of its $N_1$ neighbors. The selection matrix $\mathbf{C}$ can be written as
\begin{equation}\label{E:observ_matrix_homog_nodes}
\mathbf{C} \!=\!
\left[ \!\! \begin{array}{c c}
\mathbf{E}_{L_1+1}^T  &  \mathbf{0}_{L_1+1\times (N^2-N))}\\
\mathbf{0}_{N_1L_1\times N} & \mathbf{I}_{N_1} \otimes \mathbf{E}_{L_1}^T  \quad  \mathbf{0}_{N_1L_1\times (N^2-NN_1-N)}
\end{array} \!\! \right]\!\! .
\end{equation}
Matrix $\mathbf{C}$ has $1+L_1(1+N_1)$ rows, one per observation. The first $1+L_1$ rows correspond to the samples at node $i\!=\!1$ and the remaining $L_1N_1$ to the samples at its neighbors. Note also that matrix $\mathbf{C} \left(\mathbf{I}\otimes(\boldsymbol{\Psi} \mathbf{E}_K)\right) \bar{\boldsymbol{\Upsilon}}$ is not full (row) rank. The reason is that all the samples obtained at node $i\!=\!1$, except for the first one, are linear combinations of the samples at its neighbors. This implies that the number of frequencies that can be recovered using \eqref{E:observ_matrix_homog_nodes} is, at most, $1+L_1N_1$.

Structured observation models different from the one in \eqref{E:observ_matrix_homog_nodes} can be also of interest. For example, one can consider setups where nodes from different parts of the graph take a few samples each and forward those samples to a central fusion center. In such a case, since the nodes gathering data need not be neighbors, the problem associated with some of the samples being a linear combination of the others will not necessarily be present.

\begin{figure}[t]
\centering
\begin{subfigure}{0.33\textwidth}
  \centering
  \includegraphics[width=0.81\textwidth]{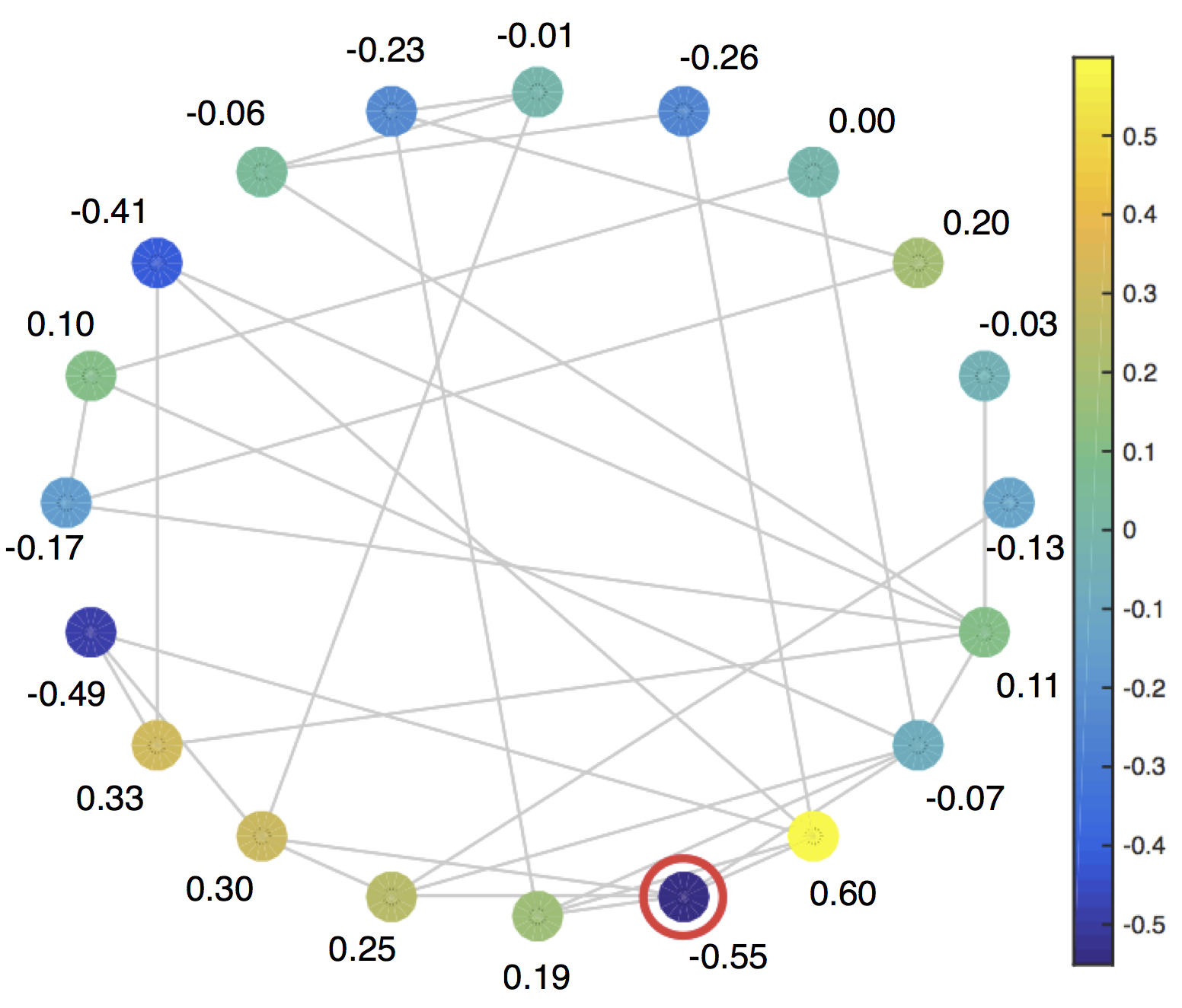}
  \caption{}
  \label{F:SyntheticSignal_graph}
\end{subfigure}%
\begin{subfigure}{0.14\textwidth}
  \centering
  \includegraphics[width=0.7\textwidth, height=4.3cm]{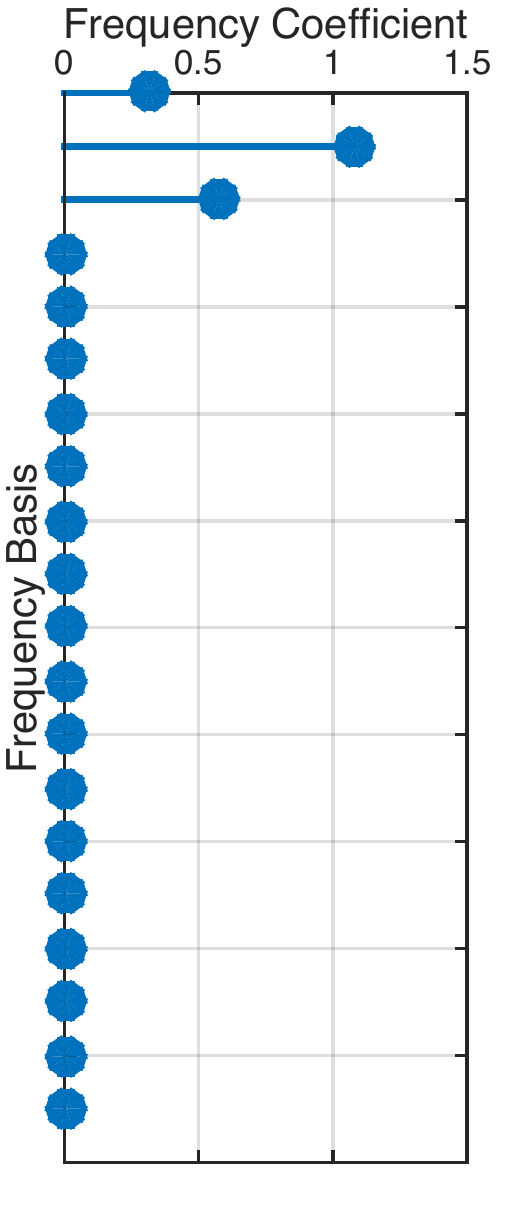}
  \caption{}
  \label{F:SyntheticSignal_freq}
\end{subfigure}
\vspace{-0.05in}
\caption{A bandlimited graph signal. (a) The graph $\mathcal{G}$ and the graph signal $\mathbf{x}$ defined on the nodes of $\mathcal{G}$. The sampling node is circled in red. (b) Frequency components $\widehat{\mathbf{x}}$ of the graph signal $\mathbf{x}$. Given that there are three non-zero coefficients, the bandwidth of signal $\mathbf{x}$ is 3.}
\vspace{-0.05in}
\label{F:SyntheticSignal}
\end{figure}

\section{Numerical experiments}\label{S:NumExper}

We start by illustrating the perfect recovery of synthetic noiseless graph signals, both when the frequency support is known and when it is not (Section~\ref{Ss:noiseless_recovery_experiments}). We then present results for real-world graph signals corresponding to the exchange among the different sectors of the economy of the United States. These are used to test recovery under the presence of noise (Section~\ref{Ss:recovery_presence_noise}) as well as to illustrate the space-shift sampling method (Section~\ref{Ss:Simus_shift_and_space}).

\subsection{Noiseless recovery and support selection}\label{Ss:noiseless_recovery_experiments}

Consider the 20-node undirected graph $\mathcal{G}$ depicted in Fig.~\ref{F:SyntheticSignal_graph}, which corresponds to a realization of a symmetric Erd\~os-–R\'enyi graph with edge probability $0.20$ \cite{bollobas1998random}. With $\mathbf{A}=\mathbf{V}\boldsymbol{\Lambda}_A\mathbf{V}^H$ denoting the adjacency matrix of $\mathcal{G}$, three different graph-shift operators are considered: $\mathbf{S}_1 = \mathbf{A}$, $\mathbf{S}_2 = \mathbf{I} - \mathbf{A}$, and $\mathbf{S}_3 = 0.5 \mathbf{A}^2$. Notice that, even though the support of $\mathbf{S}_3$ differs from that of $\mathbf{S}_1$ and $\mathbf{S}_2$, the graph-shift operator $\mathbf{S}_3$ still preserves the notion of locality as defined by a two-hop neighborhood. Note also that the three shift operators share the same set of eigenvectors $\mathbf{V}$, but they have a different set of eigenvalues.

Let $\mathbf{x}$ be a graph signal supported on $\mathcal{G}$. This signal is represented in Fig.~\ref{F:SyntheticSignal_graph}. To facilitate interpretation, the value of the signal at a given node is written explicitly next to the node and also coded by the color of the node. Although seemingly random in the node domain, the structure of the signal $\mathbf{x}$ is highly determined by the graph. To illustrate this, Fig.~\ref{F:SyntheticSignal_freq} presents the frequency components $\widehat{\mathbf{x}}$ of signal $\mathbf{x}$, where the graph frequency basis are given by the columns of $\mathbf{V}$. The figure reveals that $\mathbf{x}$ has a bandwidth of $K=3$. Since $\mathbf{V}$ is the basis for $\mathbf{S}_1$, $\mathbf{S}_2$ and $\mathbf{S}_3$, the frequency representation $\widehat{\mathbf{x}}$ and the bandwidth $K$ are the same for any of the three operators. As a result, the procedure described in Section~\ref{S:sampling_GS_local_aggregat} allows recovering the whole signal using three aggregated samples, no mater which operator is chosen for the aggregation.

\begin{figure}[t]
\centering
\includegraphics[width=0.38\textwidth]{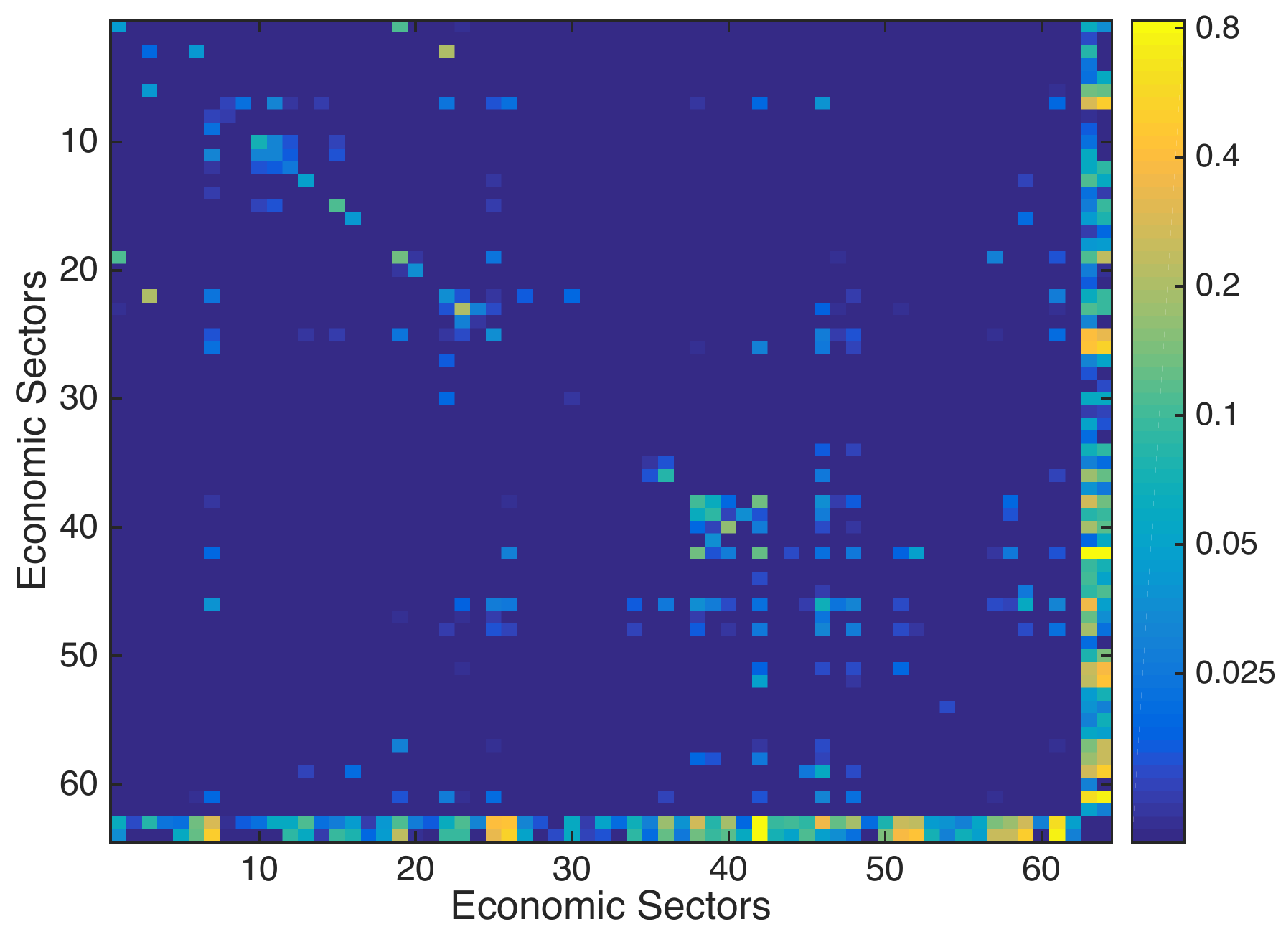}
\vspace{-0.09in}
\caption{Heat map of the graph-shift operator $\bbS$ of the economic network. It is sparse across the real economic sectors (from sector 1 to 62) while the artificial sectors AV and FU are highly connected.}
\vspace{-0.05in}
\label{F:network_imagesc_log}
\end{figure}

\begin{figure*}
\centering

\begin{subfigure}{.31\textwidth}
  \centering
  \includegraphics[width=\textwidth]{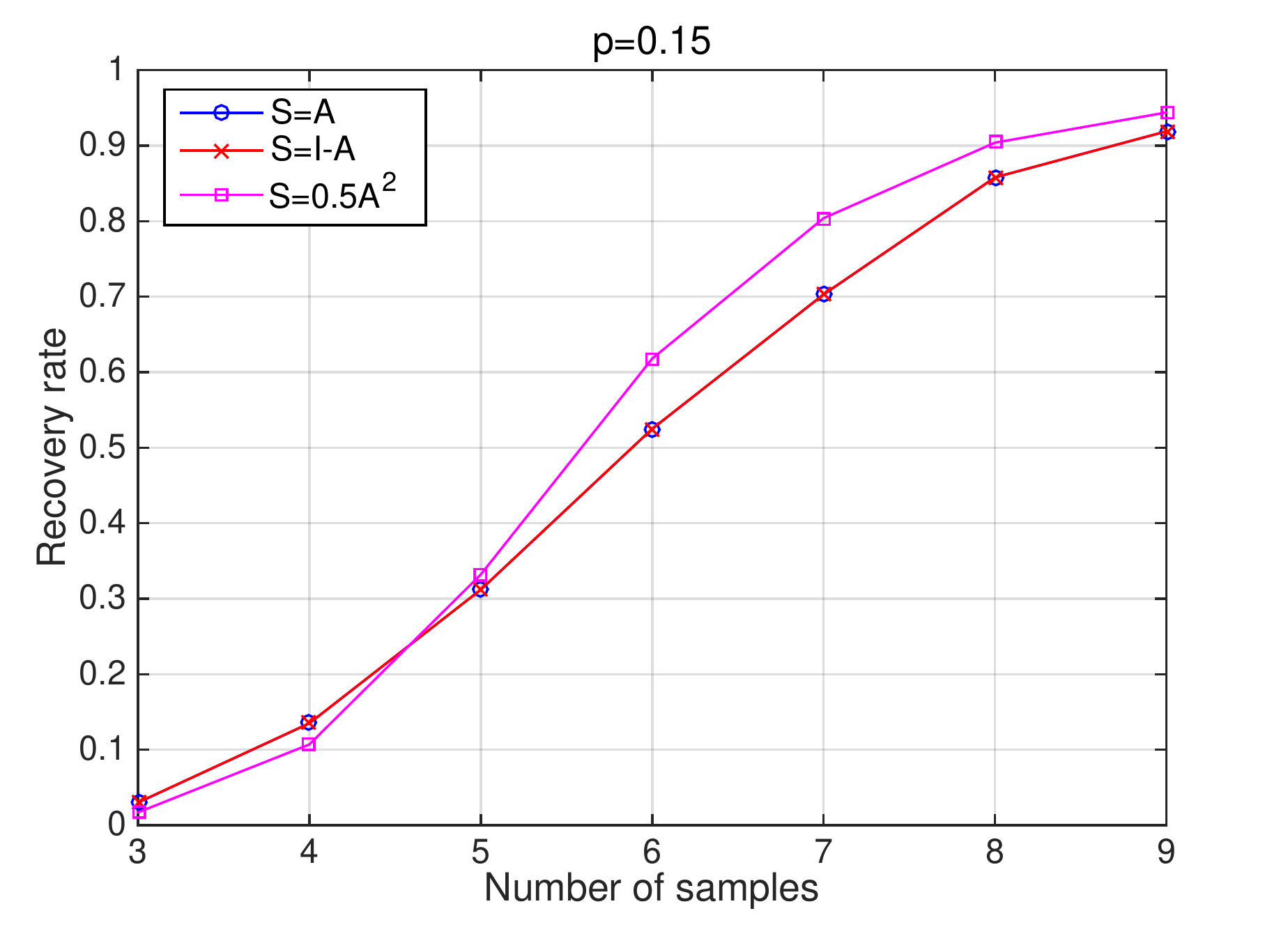}
  \caption{}
  \label{fig:unk_supp_sub1}
\end{subfigure}%
\begin{subfigure}{.31\textwidth}
  \centering
  \includegraphics[width=\textwidth]{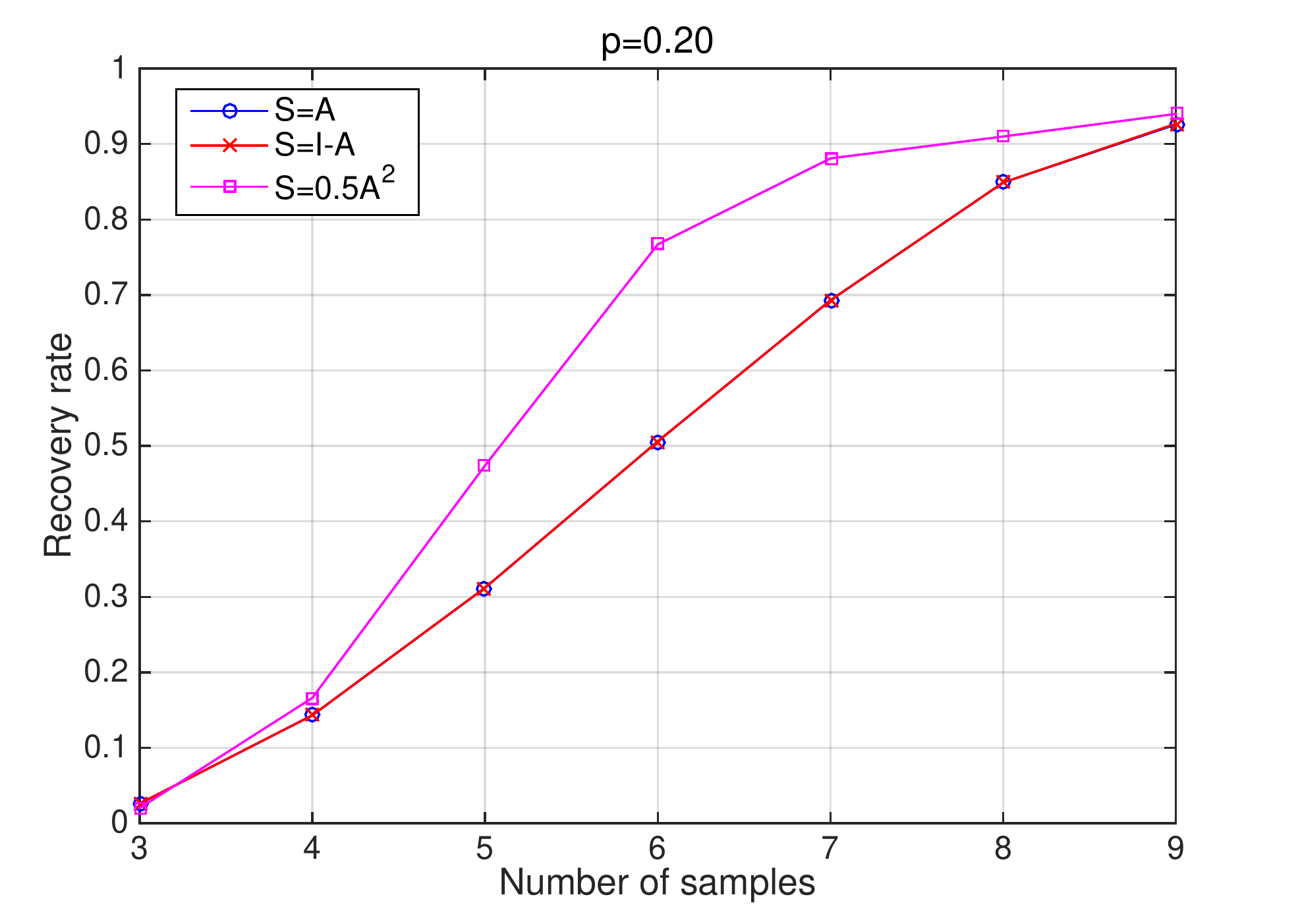}
  \caption{}
  \label{fig:unk_supp_sub2}
\end{subfigure}%
\begin{subfigure}{.31\textwidth}
  \centering
  \includegraphics[width=\textwidth]{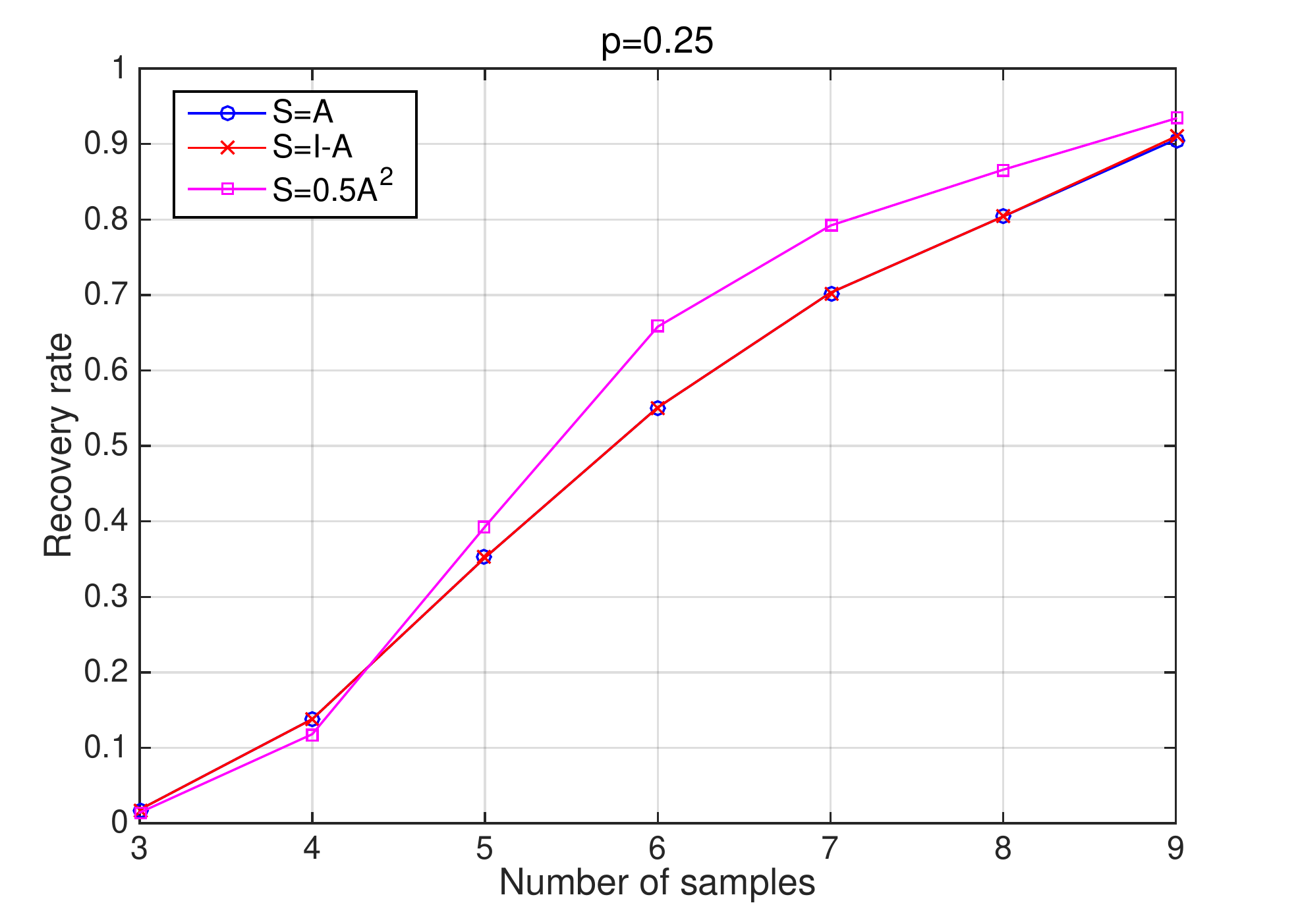}
  \caption{}
  \label{fig:unk_supp_sub3}
\end{subfigure}

\vspace{-0.05in}
\caption{Recovery rate of bandlimited signals in random graphs when the frequency support is unknown. Signals are recovered through the norm-1 relaxation of problem \eqref{E:sampled_shifted_signal_unkown} for different number of observations $K$ and three different graph-shift operators $\mathbf{S}_1 = \mathbf{A}$ (blue), $\mathbf{S}_2 = \mathbf{I} - \mathbf{A}$ (red), and $\mathbf{S}_3 = 0.5 \mathbf{A}^2$ (magenta). Random graphs with different edge probabilities were considered: (a) 0.15 , (b) 0.20, and (c) 0.25. The shift operator $\mathbf{S}_3$ consistently outperforms the others, which can be attributed to a lower matrix coherence.}
\vspace{-0.05in}
 \label{F:UnknownRecovNorm1}
\end{figure*}

Suppose that we select node $i=4$ as sampling node, which is circled in red in Fig.~\ref{F:SyntheticSignal_graph}. If the shift is $\mathbf{S}_1$, the $3$ first observations taken by that node are $\mathbf{y}_4=[-0.55, 1.27, -2.94]^T$. The first observation corresponds to the value of the signal at node 4, the second one to the aggregated signal at its neighbors and the third observation corresponds to a linear combination of the signal values within its two-hop neighborhood. Since $K=3$, Proposition~\ref{P:recoveringcond_oursampling} guarantees recovery if: i) the $3$ first eigenvalues of the shift operator are distinct and ii) the $3$ first values of $\boldsymbol{\upsilon}_4$ are non-zero. It turns out that for $\mathbf{S}_1$ and node 4 these two conditions hold true and, hence, the interpolation in \eqref{E:interp_oursampling} yields the original signal in Fig.~\ref{F:SyntheticSignal_graph}. In fact, for the network at hand, these two conditions are satisfied for all nodes and shift operators considered. This implies that perfect reconstruction in a noiseless setting is achieved independently of which node aggregates the information and which shift operator -- among the three presented -- is picked. To better asses the conditions in Proposition~\ref{P:recoveringcond_oursampling}, we build 10,000 different random graphs where the edge probability is randomly chosen from the interval $[0.15, 0.25]$. Realizations that do not give rise to a connected graph are discarded. We vary the number of nodes from 10 to 30 and the active frequencies of the graph signals from 1 to 5. For each random graph and signal defined on it, we test for perfect signal recovery on every node. The simulations show that in $99.89\%$ of the cases the signal is successfully recovered.

The graph signal $\mathbf{x}$ can be sampled and recovered even when the frequency support is unknown, i.e., when we know that $\widehat{\mathbf{x}} = \mathbf{V}^{-1} \mathbf{x}$ contains $K=3$ non-zero components, but we do not know the indices of these $K$ active frequencies. In this case, however, $2K=6$ samples are needed to guarantee identifiability (cf. Proposition~\ref{P:joint_recovery_supportid}). By solving problem \eqref{E:sampled_shifted_signal_unkown}, the signal can be recovered at every node and using any of the three shift operators, as in the previous case. However, when solving a relaxed version of problem \eqref{E:sampled_shifted_signal_unkown}, accurate signal recovery depends on the specific network, signal and node selected for reconstruction. Moreover, the recovery rate depends on the choice of the graph-shift operator $\mathbf{S}$. For example, for the signal in Fig.~\ref{F:SyntheticSignal_graph}, solving a 1-norm relaxation of the problem \eqref{E:sampled_shifted_signal_unkown} yields the original graph signal $\mathbf{x}$ if $\mathbf{S}=\mathbf{A}$ and $i=4$, but fails if $\mathbf{S}= \mathbf{I} - \mathbf{A}$ and $i=5$ where node $i=5$ is the right neighbor of node $i=4$. To assess recovery better, Fig.~\ref{F:UnknownRecovNorm1} plots the success rate -- fraction of realizations for which the actual signal was recovered -- for graph-shifts $\mathbf{S}_1$, $\mathbf{S}_2$ and $\mathbf{S}_3$, and different number of observations. Each point in the plots represents an average across all nodes in the network, 5 signal realizations and 10 random graph realizations. The three plots correspond to symmetric Erd\~os-–R\'enyi graphs generated using different edge probabilities: 0.15, 0.20, and 0.25. The recovery rate for $\mathbf{S}_3 = 0.5 \mathbf{A}^2$ is consistently higher than for the other shift operators considered. This is not surprising: when squaring the adjacency matrix to generate $\mathbf{S}_3$, the dissimilarity between any pair of eigenvalues is increased, which reduces the matrix coherence $\mu_i(\mathbf{C})$ associated with $\mathbf{S}_3 = 0.5 \mathbf{A}^2$ and facilitates sparse recovery (cf. last paragraph in Section~\ref{Ss:noiseless_joint_recovery_support_identification}). Nonetheless, if success rate is the main concern, there exist relaxations of the 0-norm that give better results than the 1-norm used \cite{cai2011orthogonal}.

\subsection{Recovery in the presence of noise}\label{Ss:recovery_presence_noise}

\begin{figure*}
\centering

\begin{subfigure}{.32\textwidth}
  \centering
  \includegraphics[width=\textwidth]{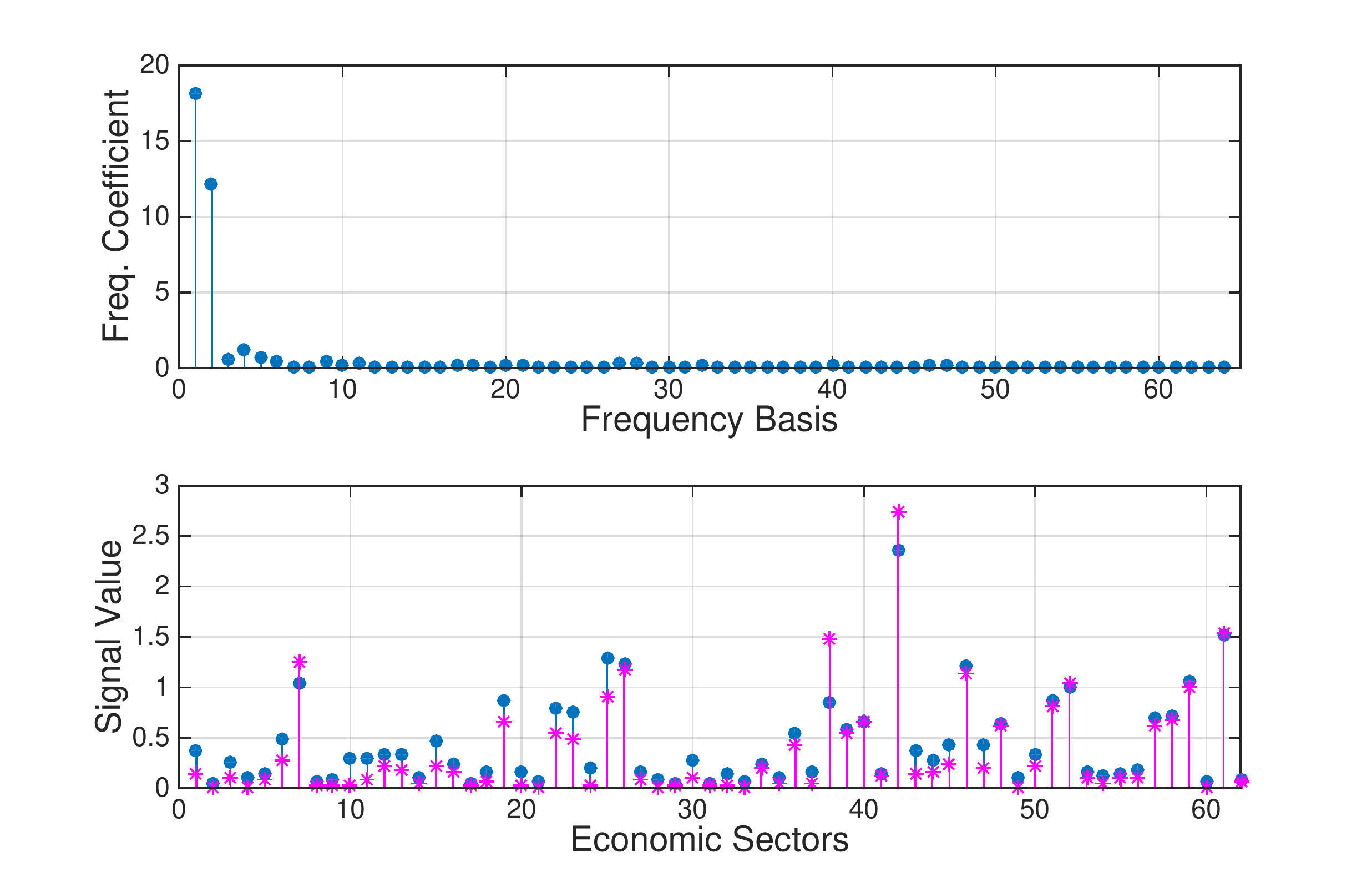}
  \caption{}
  \label{fig:sub1}
\end{subfigure}%
\begin{subfigure}{.32\textwidth}
  \centering
  \includegraphics[width=\textwidth]{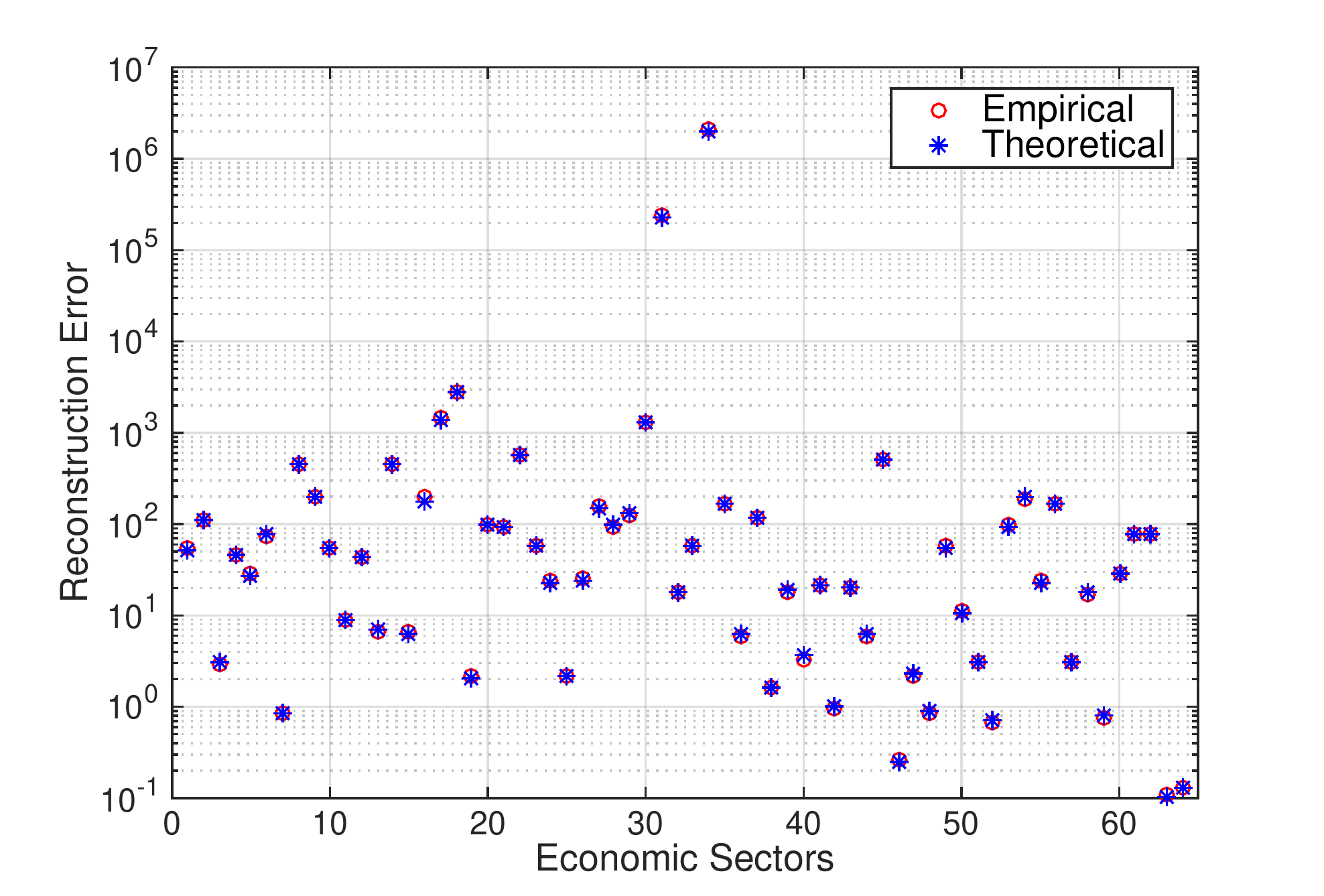}
  \caption{}
  \label{fig:sub2}
\end{subfigure}%
\begin{subfigure}{.32\textwidth}
  \centering
  \includegraphics[width=\textwidth]{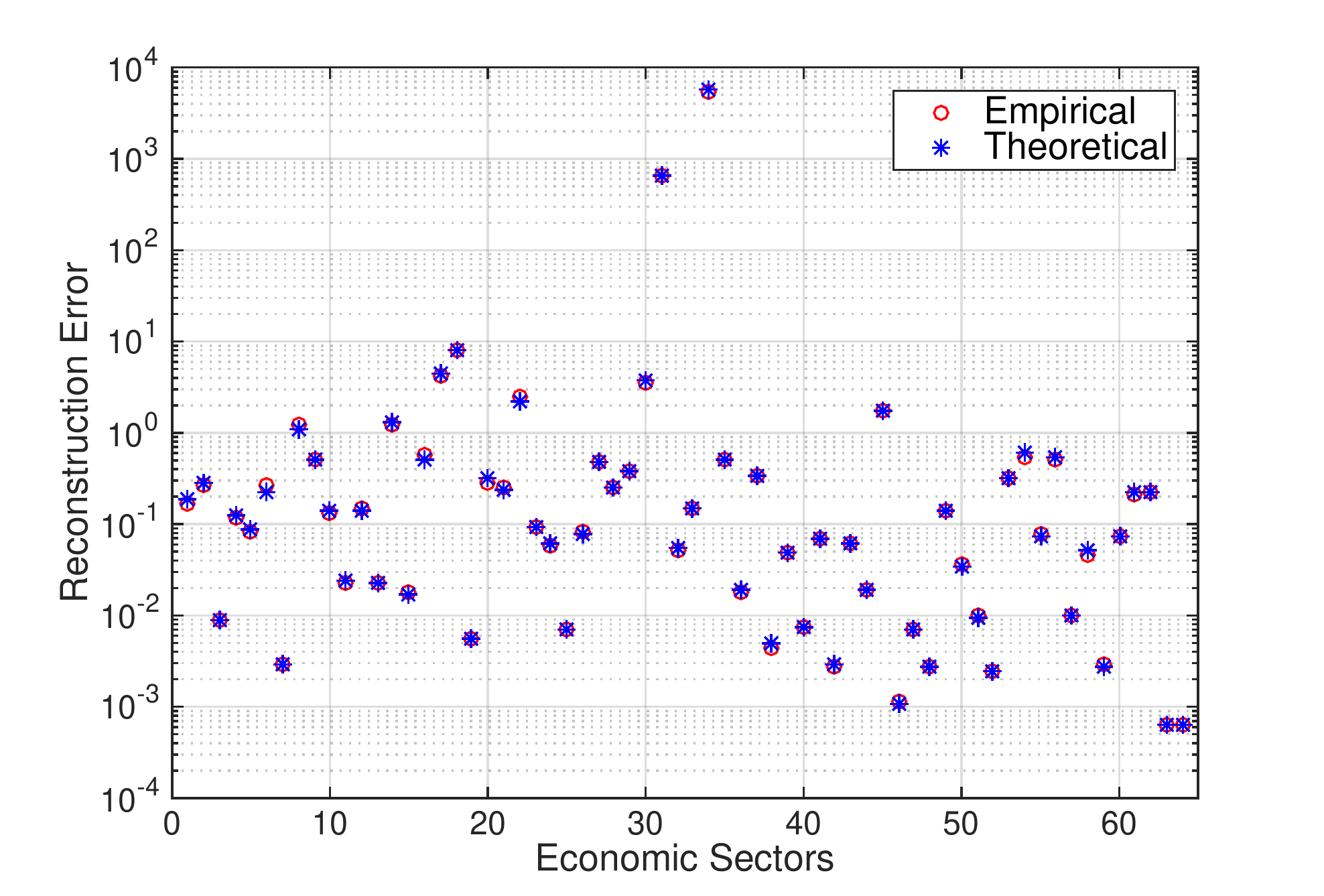}
  \caption{}
  \label{fig:sub3}
\end{subfigure}

\begin{subfigure}{.32\textwidth}
  \centering
  \includegraphics[width=\textwidth]{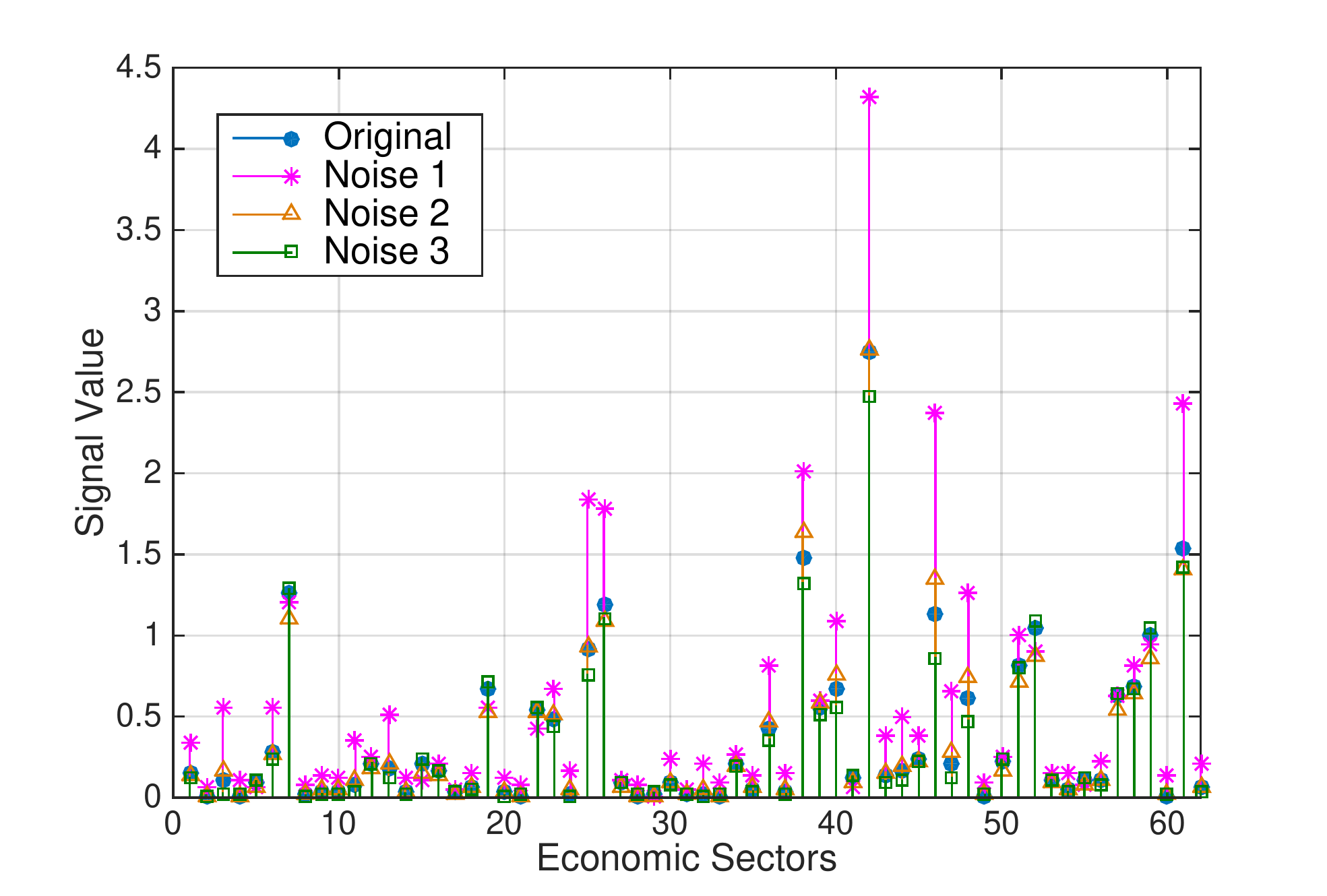}
  \caption{}
  \label{fig:sub4}
\end{subfigure}%
\begin{subfigure}{.32\textwidth}
  \centering
  \includegraphics[width=\textwidth]{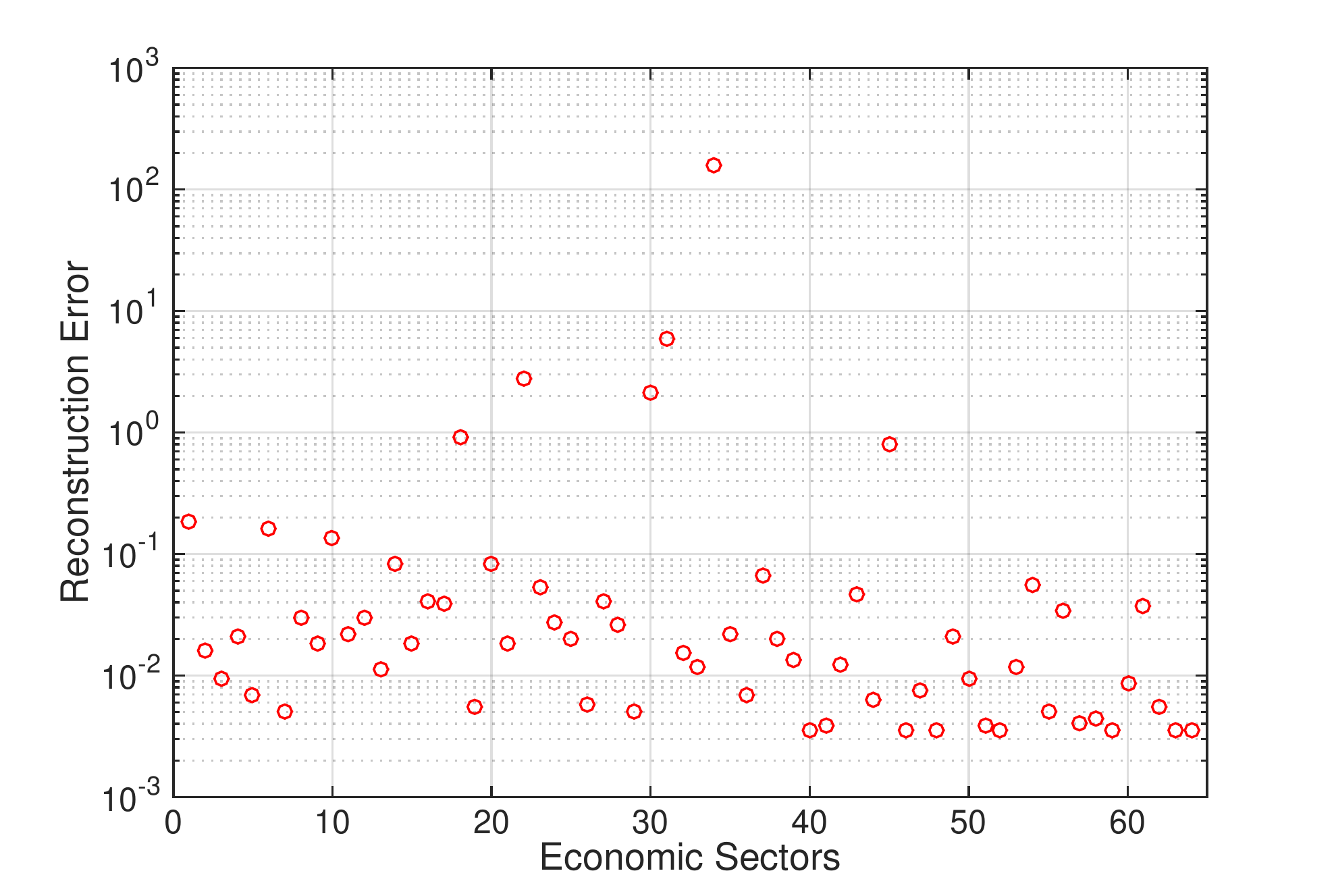}
  \caption{}
  \label{fig:sub5}
\end{subfigure}%
\begin{subfigure}{.32\textwidth}
  \centering
  \includegraphics[width=\textwidth]{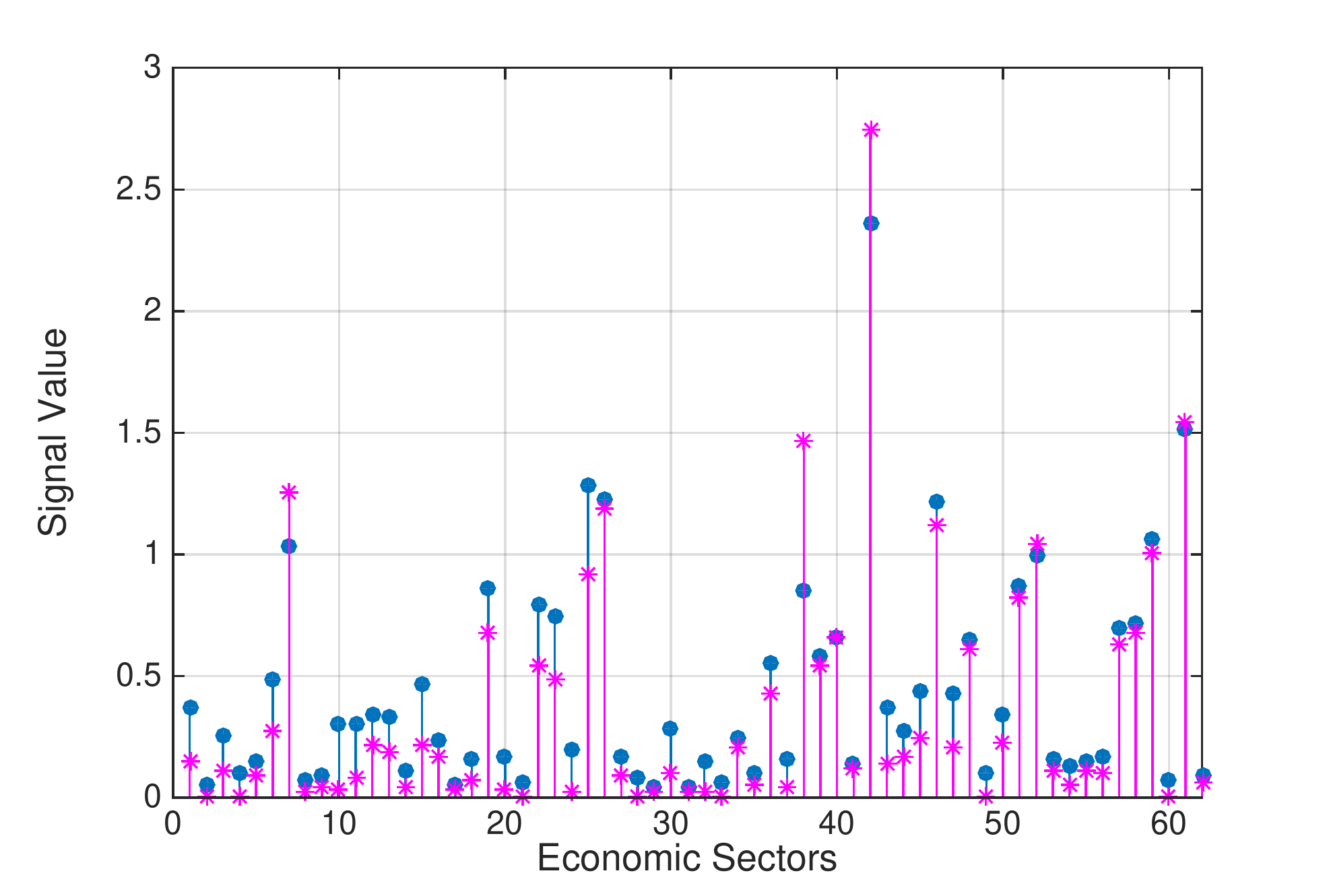}
  \caption{}
  \label{fig:sub6}
\end{subfigure}

\caption{(a) Top: Frequency representation of the graph signal $\mathbf{x}$ in the basis of eigenvectors of the graph-shift $\mathbf{S}$. The signal is approximately bandlimited. Bottom: Signal $\mathbf{x}$ (blue) and its reconstruction $\mathbf{x}_4$ (magenta) when keeping only the first 4 frequency components. (b) Empirical (red circle) and theoretical (blue star) reconstruction errors for different sampling nodes when white noise is added to the observed signal. (c) Empirical (red circle) and theoretical (blue star) reconstruction errors when white noise is added directly to the signal $\mathbf{x}_4$. (d) Signal $\mathbf{x}_4$ (blue) and the best reconstruction achieved when sampling an economic sector for the three types of noise considered: white noise in the observations (magenta), white noise in the signal (orange) and white noise in the active frequency components (green). (e) Reconstruction errors for different sampling nodes when interpolating signal $\mathbf{x}$ based on four observations. (f) Signal $\mathbf{x}$ (blue) and the best reconstruction (magenta) achieved when performing local aggregation sampling of economic sectors.}
\vspace{-0.1in}
\label{fig_experiments}
\end{figure*}

The Bureau of Economic Analysis of the U.S. Department of Commerce publishes a yearly table of input and outputs organized by economic sectors \cite{USinputoutput}. More precisely, we have a set $\mathcal{N}$ of 62 industrial sectors as defined by the North American Industry Classification System and a similarity function $U: \mathcal{N} \times \mathcal{N} \to \reals_+$ where $U(i, i')$ represents how much of the production of sector $i$, expressed in trillions of dollars per year, was used as an input of sector $i'$ on average during years 2008, 2009, and 2010. Moreover, for each sector we are given two economic markers: the added value (AV) generated and the level of production destined to the market of final users (FU). Thus, we define a graph on the set of $N=64$ nodes comprising the original 62 sectors plus the two artificial ones (AV and FU) and an associated symmetric graph-shift operator $\bar{\mathbf{S}}$ defined as $\bar{\mathbf{S}}_{ij} = (U(i, j)+U(j,i))/2$. We then threshold $\bar{\mathbf{S}}$ in order to increase its sparsity by setting to 0 all the values lower than 0.01 to obtain the shift operator $\mathbf{S}=\mathbf{V}\boldsymbol{\Lambda}\mathbf{V}^{H}$, which is normal given that it is symmetric; see Fig.~\ref{F:network_imagesc_log}. Consider the signal $\mathbf{x} \in \reals^{64}$ on the mentioned graph where $\mathbf{x}$ contains the total production -- in trillion of dollars -- of each sector (including AV and FU) during year 2011. Signal $\mathbf{x}$ is approximately bandlimited in $\mathbf{S}$ since most of the elements of $\widehat{\mathbf{x}} = \mathbf{V}^{H} \mathbf{x}$ are close to zero; see Fig.~\ref{fig:sub1} (top). In particular, the reconstructed signal $\mathbf{x}_4 = \mathbf{V}_4 \widehat{\mathbf{x}}_4$ obtained by just keeping the first four frequency coefficients attains a reconstruction error of $3.5 \times 10^{-3}$  computed as the ratio between the energy of the error and the energy of the original signal. This small reconstruction error is nonetheless noticeable when plotting the original signal $\mathbf{x}$ and the reconstructed one $\mathbf{x}_4$; see Fig.~\ref{fig:sub1} (bottom). To present a reasonable scale for illustration, sectors AV and FU are not included in Fig.~\ref{fig_experiments}, since the signal takes out-of-scale values for these sectors.

In Sections~\ref{Sss:synthetic_observed_signal} to \ref{Sss:synthetic_active_frequencies} we consider the bandlimited signal $\mathbf{x}_4$ as noiseless and add different types of Gaussian noise to analyze the interpolation performance at different nodes. Differently, in Section~\ref{Sss:real_world_noisy_signal} we interpret the whole graph signal $\mathbf{x}$ as a noisy version of $\mathbf{x}_4$ and analyze the reconstruction error when interpolating $\mathbf{x}$ from just 4 samples.

\subsubsection{White noise in the observed signal}\label{Sss:synthetic_observed_signal}
We perform aggregation sampling of multiple noisy versions of $\mathbf{x}_4$ via successive applications of the graph-shift $\mathbf{S}$ at different economic sectors (nodes). The noisy versions of $\mathbf{x}_4$ are generated by adding noise to the observed signal as described in \eqref{E:noise_model_1}. The power of the white noise $\sigma^2$ is the same for all nodes and is computed so that, averaging across nodes, the linear signal to noise ratio (SNR) for the first, second, third and fourth observations in each node is 2, 10, 50, and 250, respectively. This increase in SNR is attributable to the fact that successive applications of the shift $\mathbf{S}$ increase the signal magnitude while we keep the noise power $\sigma^2$ constant. In Fig.~\ref{fig:sub2} we plot the empirical average reconstruction error at different nodes across 1,000 noisy realizations of $\mathbf{x}_4$ and compare it with the theoretical average error, i.e., the trace of $\mathbf{R}_e^{(i)}$ in \eqref{E:cov_error_est_time_noise} [cf.~\eqref{E:error_metric_trace}]. We first observe that the computed theoretical error indeed coincides with the average empirical error across realizations. Moreover, notice that the reconstruction performance is highly node dependent. The error is minimized for the reconstruction based on the artificial sectors AV and FU. This is reasonable since these two nodes -- unlike other sectors -- are closely related to every other sector of the economy (cf. Fig.~\ref{F:network_imagesc_log}). Furthermore, the sectors achieving the worst reconstruction errors are `Publishing Industries' and `Ground Passenger Transportation' corresponding to nodes 34 and 31. This can be explained by analyzing the vectors $\bar{\boldsymbol{\upsilon}}_{34}  = \boldsymbol{\upsilon}_{34} \mathbf{E}_4$ and $\bar{\boldsymbol{\upsilon}}_{31}$ (cf. Lemma~\ref{L:vandermonde_matrix}). Even though both vectors have all four components different from zero, which guarantees perfect reconstruction in the noiseless case (cf. Proposition~\ref{P:recoveringcond_oursampling}), they possess an element whose absolute value is in the order of $10^{-4}$, increasing the sensitivity of the reconstruction in the presence of noise. For all other nodes the smallest element of $\bar{\boldsymbol{\upsilon}}_{i}$ is at least one order of magnitude larger. Fig.~\ref{fig:sub4} presents the reconstruction obtained by aggregation sampling in node 46 corresponding to `Professional Services' -- best among real economic sectors, i.e., excluding AV and FU -- which achieves an error of 0.26.

\subsubsection{White noise in the original signal}\label{Sss:synthetic_original_signal}
Similarly to the analysis performed in the previous section, we investigate the reconstruction performance of aggregation sampling at different nodes. However, in this case, the noise is added to the original signal, following the model described in \eqref{E:noise_model_2}. The power of the white noise $\sigma^2$ is set to induce a linear SNR of $10^2$. As was the case in the previous section, the average empirical error (across 1,000 realizations) matches closely our theoretical estimates; see Fig.~\ref{fig:sub3}. Moreover, the specific nodes that lead to a good (bad) interpolation performance are very similar to those in the previous noise model. Indeed, sectors 34 and 31 have the highest reconstruction error whereas AV and FU attain the best reconstructions. Fig.~\ref{fig:sub4} shows the best reconstruction -- excluding AV and FU -- which amounts to an error of 0.001 and corresponds to the sector `Professional Services' at node 46.

\subsubsection{White noise in the active frequencies}\label{Sss:synthetic_active_frequencies}
We consider a third category of noisy versions of $\mathbf{x}_4$ obtained by adding white noise only to the four active frequencies, as described in \eqref{E:noise_model_3}. The power of the white noise $\sigma^2$ is set to induce a linear SNR of $10^2$. The empirical reconstruction error associated with each node -- averaged over 1,000 noisy realizations of $\mathbf{x}_4$ -- is the same among nodes. This validates the analysis in \eqref{E:node_choice_noise_3}, which stated that, for this noise model, the quality of the reconstruction is node independent. In Fig.~\ref{fig:sub4} we present an example of such reconstruction, achieving an error of 0.01.

\subsubsection{Real-world noisy signal}\label{Sss:real_world_noisy_signal}
We interpret the graph signal $\mathbf{x}$ as a noisy realization of a signal of bandwidth 4. Hence, our goal is to obtain the best reconstruction of $\mathbf{x}$ based on 4 observations. As described in \eqref{E:cov_error_est_time_noise} and shown before, interpolation performance is highly node dependent. Indeed, the reconstruction error when keeping the first 4 observations at each node spans 5 orders of magnitude depending on the sampling node, although for most nodes it is contained between $10^{-3}$ and $10^{-1}$; see Fig.~\ref{fig:sub5}. The best reconstruction among the real sectors is achieved by `Insurance Carriers' (node 40). The best and the median reconstructions are acceptable, attaining errors of 0.0035 and 0.019, respectively. Fig.~\ref{fig:sub6} depicts the best reconstruction.

\subsection{Space-shift sampling}\label{Ss:Simus_shift_and_space}

In Section~\ref{Sss:real_world_noisy_signal} we analyzed the accuracy of interpolating the U.S. economic activity after aggregation sampling in different economic sectors. The minimum and median reconstruction errors are presented in the first row of Table~\ref{tab_space_shift_errors}, where the reconstruction error is quantified as the ratio between the energy of the error and that of the original signal. An alternative approach is to implement selection sampling, i.e. to sample the signal $\mathbf{x}$ in 4 different sectors -- excluding the artificial sectors AV and FU -- and interpolate the whole signal from these 4 observations, as explained in Section~\ref{Ss:conventional sampling}. Recall that reconstruction is not guaranteed for every subset of 4 nodes since we must have invertibility of $(\mathbf{C}\mathbf{V}_K)$ [cf. \eqref{E:interp_regsampling}]. By analyzing the minimum and median reconstruction errors -- see the two first rows in Table~\ref{tab_space_shift_errors} -- it is clear that the node aggregation sampling outperforms the node selection sampling. This is intuitive since most of the energy of the signal is contained in the two first frequencies [cf. Fig.~\ref{fig:sub1}(top)], which are associated with the largest eigenvalues. Hence, after successive implementations of the graph-shift, the error in estimating these frequencies is reduced, resulting in a smaller error in the interpolation of the whole signal.

As developed in Section~\ref{S:shift_space_bigvec_sampling}, more general sampling strategies can be implemented. For example, we can sample the value of the signal at 4 nodes after the application of one, two or three graph-shifts. The results -- listed in rows 3, 4 and 5 of Table~\ref{tab_space_shift_errors}-- reveal that reduction in the median error after each graph-shift application is conspicuous, especially when going from no applications -- median error of 4.2 -- to one application -- median error of 0.03. A different alternative is a sampling strategy that selects the original signal and the signal after one shift in two different sectors. The results, listed in the last row of Table~\ref{tab_space_shift_errors}, show that this configuration leads to a very good reconstruction performance: 0.0035 minimum error and 0.039 median error. Note that with this sampling configuration, the two sectors are only required to compute the aggregated activity of  their one-hop neighbors.

\begin{table}[t]
\centering
\renewcommand{\arraystretch}{1.2}
\begin{tabular}{c c c c c c}
\hline
\multicolumn{4}{c}{Sampling strategy} & Min. error & Median error \\\hline
$[\mathbf{x}]_i$  & $[\mathbf{Sx}]_i$ & $[\mathbf{S^2x}]_i$ & $[\mathbf{S^3x}]_i$ & .0035 & .019 \\
$[\mathbf{x}]_i$  & $[\mathbf{x}]_j$& $[\mathbf{x}]_k$& $[\mathbf{x}]_l$ & .0039 & 4.2 \\
$[\mathbf{Sx}]_i$  & $[\mathbf{Sx}]_j$& $[\mathbf{Sx}]_k$& $[\mathbf{Sx}]_l$ & .0035 & .030 \\
$[\mathbf{S^2x}]_i$  & $[\mathbf{S^2x}]_j$& $[\mathbf{S^2x}]_k$& $[\mathbf{S^2x}]_l$ & .0035 & .0055 \\
$[\mathbf{S^3x}]_i$  & $[\mathbf{S^3x}]_j$& $[\mathbf{S^3x}]_k$& $[\mathbf{S^3x}]_l$ & .0035 & .0040 \\
$[\mathbf{x}]_i$  & $[\mathbf{Sx}]_i$& $[\mathbf{x}]_j$& $[\mathbf{Sx}]_j$ & .0035 & .039 \\
\hline
\end{tabular}
\caption{Minimum and median reconstruction error -- energy of the error as a fraction of the energy of the signal $\mathbf{x}$ -- for different sampling strategies. The first sampling strategy corresponds to aggregation sampling, i.e., observing the same node $i$ after successive applications of 0, 1, 2, and 3 graph-shifts $\mathbf{S}$. The second sampling strategy corresponds to selection sampling, i.e., observing the value of the signal $\mathbf{x}$ at 4 different nodes $i,j,k,l$. The remaining strategies correspond to more general space-shift sampling schemes.}
\label{tab_space_shift_errors}
\end{table}

\section{Conclusions}\label{S:Concl}
A novel scheme for sampling bandlimited graph signals -- that admit a sparse representation in the frequency domain -- was proposed.
The scheme was based on the aggregation of local information at a single node after successive applications of the graph-shift operator. This contrasted to most existing works, which focus on sampling the value of the signal observed at a subset of nodes.
Our scheme was shown to be equivalent to classical sampling for directed cycle graphs whereas, for more general graphs, the Vandermonde structure of the sampling matrix was exploited to determine the conditions for perfect reconstruction in the absence of noise. Reconstruction under correlated noise was analyzed, and design criteria to pick the sampling node and shifts leading to optimal noisy reconstruction were discussed. Scenarios where the specific set of frequencies present in the bandlimited signal is not known were also investigated and connections with sparse signal reconstruction were drawn.
Finally, a more general sampling scheme was presented which contained, as particular cases, the selection sampling as well as our local aggregation approach.
The various sampling and interpolation scenarios were illustrated through numerical experiments in both synthetic and real-world graph signals.

\bibliographystyle{IEEEtran}
%
\bibliography{citations}

\begin{thebibliography}{10}
\providecommand{\url}[1]{#1}
\csname url@samestyle\endcsname
\providecommand{\newblock}{\relax}
\providecommand{\bibinfo}[2]{#2}
\providecommand{\BIBentrySTDinterwordspacing}{\spaceskip=0pt\relax}
\providecommand{\BIBentryALTinterwordstretchfactor}{4}
\providecommand{\BIBentryALTinterwordspacing}{\spaceskip=\fontdimen2\font plus
\BIBentryALTinterwordstretchfactor\fontdimen3\font minus
  \fontdimen4\font\relax}
\providecommand{\BIBforeignlanguage}[2]{{%
\expandafter\ifx\csname l@#1\endcsname\relax
\typeout{** WARNING: IEEEtran.bst: No hyphenation pattern has been}%
\typeout{** loaded for the language `#1'. Using the pattern for}%
\typeout{** the default language instead.}%
\else
\language=\csname l@#1\endcsname
\fi
#2}}
\providecommand{\BIBdecl}{\relax}
\BIBdecl

\bibitem{unser2000sampling}
M.~Unser, ``Sampling-50 years after {S}hannon,'' \emph{Proc. IEEE}, vol.~88,
  no.~4, pp. 569--587, 2000.

\bibitem{EmergingFieldGSP}
D.~I. Shuman, S.~K. Narang, P.~Frossard, A.~Ortega, and P.~Vandergheynst, ``The
  emerging field of signal processing on graphs: Extending high-dimensional
  data analysis to networks and other irregular domains,'' \emph{IEEE Signal
  Process. Mag.}, vol.~30, no.~3, pp. 83--98, 2013.

\bibitem{SandryMouraSPG_TSP13}
A.~Sandryhaila and J.~Moura, ``Discrete signal processing on graphs,''
  \emph{IEEE Trans. Signal Process.}, vol.~61, no.~7, pp. 1644--1656, April
  2013.

\bibitem{RabICASSP12_ApproxSignalsGraphs}
X.~Zhu and M.~Rabbat, ``Approximating signals supported on graphs,'' in
  \emph{IEEE Intl. Conf. Acoust., Speech and Signal Process. (ICASSP)}, March
  2012, pp. 3921--3924.

\bibitem{SamplingOrtegaICASSP14}
A.~Anis, A.~Gadde, and A.~Ortega, ``Towards a sampling theorem for signals on
  arbitrary graphs,'' in \emph{IEEE Intl. Conf. Acoust., Speech and Signal
  Process. (ICASSP)}, 2014, pp. 3864--3868.

\bibitem{AlgFindSupportSamplGlobalsip2014}
\BIBentryALTinterwordspacing
I.~Shomorony and A.~S. Avestimehr, ``Sampling large data on graphs,''
  \emph{CoRR}, vol. abs/1411.3017, 2014. [Online]. Available:
  \url{http://arxiv.org/abs/1411.3017}
\BIBentrySTDinterwordspacing

\bibitem{SamplingKovacevicMoura_1415}
S.~Chen, A.~Sandryhaila, J.~M. Moura, and J.~Kova{\v{c}}evi{\'c}, ``Signal
  recovery on graphs,'' \emph{arXiv preprint arXiv:1411.7414}, 2014.

\bibitem{chen2015discrete}
S.~Chen, R.~Varma, A.~Sandryhaila, and J.~Kova{\v{c}}evi{\'c}, ``Discrete
  signal processing on graphs: Sampling theory,'' \emph{arXiv preprint
  arXiv:1503.05432}, 2015.

\bibitem{wang2014local}
X.~Wang, P.~Liu, and Y.~Gu, ``Local-set-based graph signal reconstruction,''
  \emph{arXiv preprint arXiv:1410.3944}, 2014.

\bibitem{donoho2003spark}
D.~L. Donoho and M.~Elad, ``Optimally sparse representation in general
  (nonorthogonal) dictionaries via $\ell^1$ minimization,'' \emph{Proc. Nat.
  Academy of Sciences}, vol. 100, no.~5, pp. 2197--2202, 2003.

\bibitem{candes2006robust}
E.~Cand{\`e}s, J.~Romberg, and T.~Tao, ``Robust uncertainty principles: Exact
  signal reconstruction from highly incomplete frequency information,''
  \emph{IEEE Trans. Inf. Theory}, vol.~52, no.~2, pp. 489--509, 2006.

\bibitem{elad2007optimized}
M.~Elad, ``Optimized projections for compressed sensing,'' \emph{IEEE Trans.
  Signal Process.}, vol.~55, no.~12, pp. 5695--5702, 2007.

\bibitem{RabICASSP12_SpectCompressSensingGraphs}
X.~Zhu and M.~Rabbat, ``Graph spectral compressed sensing for sensor
  networks,'' in \emph{IEEE Intl. Conf. Acoust., Speech and Signal Process.
  (ICASSP)}, March 2012, pp. 2865--2868.

\bibitem{SandryMouraSPG_TSP14Freq}
A.~Sandryhaila and J.~Moura, ``Discrete signal processing on graphs: Frequency
  analysis,'' \emph{IEEE Trans. Signal Process.}, vol.~62, no.~12, pp.
  3042--3054, June 2014.

\bibitem{godsil2001algebraic}
C.~Godsil and G.~Royle, \emph{Algebraic graph theory}.\hskip 1em plus 0.5em
  minus 0.4em\relax Springer-Verlag, Graduate Texts in Mathematics, 2001, vol.
  207.

\bibitem{RabICASSP14_SpectCharSigsSmallWord}
M.~Rabbat and V.~Gripon, ``Towards a spectral characterization of signals
  supported on small-world networks,'' in \emph{IEEE Intl. Conf. Acoust.,
  Speech and Signal Process. (ICASSP)}, May 2014, pp. 4793--4797.

\bibitem{macon1958inverses}
N.~Macon and A.~Spitzbart, ``Inverses of vandermonde matrices,'' \emph{American
  Mathematical Monthly}, pp. 95--100, 1958.

\bibitem{KayBook}
S.~M. Kay, \emph{Fundamentals of Statistical Signal Processing: Estimation
  Theory}.\hskip 1em plus 0.5em minus 0.4em\relax Upper Saddle River, NJ, USA:
  Prentice-Hall, Inc., 1993.

\bibitem{pukelsheim1993optimal}
F.~Pukelsheim, \emph{Optimal design of experiments}.\hskip 1em plus 0.5em minus
  0.4em\relax SIAM, 1993, vol.~50.

\bibitem{Eusipco15}
S.~Segarra, A.~G. Marques, G.~Leus, and A.~Ribeiro, ``Interpolation of graph
  signals using shift-invariant graph filters,'' in \emph{Proc. of European
  Signal Process. Conf. (EUSIPCO) (submitted)}, Nice, France, August 2015.

\bibitem{fullsparkframes12}
B.~Alexeev, J.~Cahill, and D.~G. Mixon, ``Full spark frames,'' \emph{J. of
  Fourier Analysis and Appl.}, vol.~18, no.~6, pp. 1167--1194, 2012.

\bibitem{bollobas1998random}
B.~Bollob{\'a}s, \emph{Random graphs}.\hskip 1em plus 0.5em minus 0.4em\relax
  Springer, 1998.

\bibitem{cai2011orthogonal}
T.~T. Cai and L.~Wang, ``Orthogonal matching pursuit for sparse signal recovery
  with noise,'' \emph{IEEE Trans. Inf. Theory}, vol.~57, no.~7, pp. 4680--4688,
  2011.

\bibitem{USinputoutput}
\BIBentryALTinterwordspacing
{Bureau of Economic Analysis}, ``Input-output accounts: the use of commodities
  by industries before redefinitions,'' \emph{U.S. Dept. of Commerce}, 2011.
  [Online]. Available: \url{http://www.bea.gov/iTable/index\_industry.cfm}
\BIBentrySTDinterwordspacing

\end{thebibliography}

\end{document}